\pdfoutput=1
\documentclass[usenatbib,psfig]{mnras}

\usepackage{times}
\usepackage{url}
\usepackage{graphicx}
\usepackage{subcaption}
\usepackage[export]{adjustbox}

\usepackage{caption}
\usepackage{changepage}
\usepackage[justification=centering]{caption}

\usepackage{wrapfig}

\usepackage{dirtytalk}
\usepackage{amssymb}
\usepackage{placeins}
\usepackage{threeparttable}
\usepackage{color}
\usepackage{float}
\newcommand{\cha}[1]{\textcolor{black}{#1}}

\DeclareRobustCommand{\VAN}[3]{#3}

\usepackage[version=4]{mhchem}
\usepackage{titlesec}

\newcommand{\Msun}{\ensuremath{\textrm{M}_{\odot}}}

\newcommand{\Lsun}{\ensuremath{\textrm{L}_{\odot}}}

\newcommand{\kms}{km~s$^{-1}$} 

\newcommand{\OI}{\mbox{O\hspace{0.25em}{\sc i}}}

\newcommand{\CII}{\mbox{C\hspace{0.25em}{\sc ii}}}
\newcommand{\CIII}{\mbox{C\hspace{0.25em}{\sc iii}}}
\newcommand{\NaI}{\mbox{Na\hspace{0.25em}{\sc i}}}
\newcommand{\MgII}{\mbox{Mg\hspace{0.25em}{\sc ii}}}

\newcommand{\SII}{\mbox{S\hspace{0.25em}{\sc ii}}}

\newcommand{\SiII}{\mbox{Si\hspace{0.25em}{\sc ii}}}
\newcommand{\SiIII}{\mbox{Si\hspace{0.25em}{\sc iii}}}

\newcommand{\CaII}{\mbox{Ca\hspace{0.25em}{\sc ii}}}

\newcommand{\FeII}{\mbox{Fe\hspace{0.25em}{\sc ii}}}
\newcommand{\FeIII}{\mbox{Fe\hspace{0.25em}{\sc iii}}}
\newcommand{\CoII}{\mbox{Co\hspace{0.25em}{\sc ii}}}
\newcommand{\CoIII}{\mbox{Co\hspace{0.25em}{\sc iii}}}
\newcommand{\NiII}{\mbox{Ni\hspace{0.25em}{\sc ii}}}

\newcommand{\Fefs}{$^{56}$Fe}

\newcommand{\Nifs}{$^{56}$Ni}

\newcommand{\KE}{E$_{\rm K}$}
\newcommand{\Dm}{\ensuremath{\Delta m_{15}(B)}}

\def\gsim{\mathrel{\rlap{\lower 4pt \hbox{\hskip 1pt $\sim$}}\raise 1pt \hbox {$>$}}}
\def\lsim{\mathrel{\rlap{\lower 4pt \hbox{\hskip 1pt $\sim$}}\raise 1pt \hbox {$<$}}}
\def\gtaprx {\lower .1ex\hbox{\rlap{\raise .6ex\hbox{\hskip .3ex
	{\ifmmode{\scriptscriptstyle >}\else
		{$\scriptscriptstyle >$}\fi}}}
	\kern -.4ex{\ifmmode{\scriptscriptstyle \sim}\else
		{$\scriptscriptstyle\sim$}\fi}}}
\def\ltaprx {\lower .1ex\hbox{\rlap{\raise .6ex\hbox{\hskip .3ex
	{\ifmmode{\scriptscriptstyle <}\else
		{$\scriptscriptstyle <$}\fi}}}
	\kern -.4ex{\ifmmode{\scriptscriptstyle \sim}\else
		{$\scriptscriptstyle\sim$}\fi}}}

\usepackage{graphicx}	
\usepackage{amsmath}	

\title[SN\,1999aa]{Abundance stratification in type Ia supernovae -- VI: the peculiar slow decliner SN\,1999aa}

\author[Aouad et al]{Charles J. Aouad$^1$\thanks{E-mail:charlesaouad@aascid.ae}, 
Paolo A. Mazzali$^{1,2}$,
Stephan Hachinger$^{3}$,
Jacob Teffs$^{1}$,
Elena Pian$^{4}$,
\newauthor 
Chris Ashall$^{5}$,
Stefano Benetti$^{6}$,
Alexei V. Filippenko$^{7}$,
Masaomi Tanaka $^{8,9}$
\\
\\
$^1$ Astrophysics Research Institute, Liverpool John Moores University, 146 Brownlow Hill, Liverpool L3 5RF, UK\\
$^2$ Max-Planck Institut f\"{u}r Astrophysik, Karl-Schwarzschild-Str. 1, D-85748 Garching, Germany\\
$^3$ Leibniz Supercomputing Centre (LRZ) of the BAdW, Boltzmannstr. 1, D-85748 Garching, Germany\\
$^4$ INAF-IASF-Bo, via Gobetti, 101, I-40129 Bologna, Italy\\
$^5$ Institute for Astronomy, University of Hawai'i at Manoa, 
2680 Woodlawn Dr., Hawai'i, HI 96822, USA\\
$^6$ INAF, Osservatorio Astronomico di Padova, Vicolo dell'Osservatorio 5, I-35122 Padova, Italy\\
$^7$ Department of Astronomy, University of California, Berkeley, CA 94720-3411, USA\\
$^8$ Astronomical Institute, Tohoku University, Aoba, Sendai 980-8578, Japan\\
$^9$ Division for the Establishment of Frontier Sciences, Organization for Advanced Studies, Tohoku University, Sendai 980-8577, Japan
}

\date{Accepted 2022 July 14. Received 2022 July 13; in original form 2022 April 3}

\pubyear{2021}

\begin{document}
\label{firstpage}
\pagerange{\pageref{firstpage}--\pageref{lastpage}}
\maketitle


\begin{abstract}
The abundance distribution in the ejecta of the peculiar slowly declining Type Ia supernova (SN\,Ia) SN\,1999aa is obtained by modelling a time series of optical spectra. Similar to SN\,1991T, SN\,1999aa was characterised by early-time spectra dominated by \FeIII\ features and a weak \SiII\,6355\,\AA\ line, but it exhibited a high-velocity \CaII\,H\&K line and morphed into a spectroscopically normal SN\,Ia earlier. Three explosion models are investigated, yielding comparable fits. The innermost layers are dominated by $\sim 0.3$\,\Msun\ of neutron-rich stable Fe-group elements, mostly stable iron. Above that central region lies a \Nifs-dominated shell, extending to $v \approx 11,000$ -- $12,000$\,\kms, with mass $\sim 0.65$\,\Msun. These inner layers are therefore similar to those of normal SNe\,Ia. However, the outer layers exhibit composition peculiarities similar to those of SN\,1991T: the intermediate-mass elements shell is very thin, containing only $\sim 0.2$\,\Msun, and is sharply separated from an outer oxygen-dominated shell, which includes $\sim 0.22$\,\Msun. These results imply that burning suddenly stopped in SN\,1999aa. This is a feature SN\,1999aa shares with SN\,1991T, and explain the peculiarities of both SNe, which are quite similar in nature apart from the different luminosities. The spectroscopic path from normal to SN\,1991T-like SNe\,Ia cannot be explained solely by a temperature sequence. It also involves composition layering differences, suggesting variations in the progenitor density structure or in the explosion parameters.
\end{abstract}



\section{INTRODUCTION}

Type Ia supernovae (SNe\,Ia) are among the most luminous transients in the Universe. They are thought to be the thermonuclear explosions of carbon-oxygen (CO) white dwarfs close to the Chandrasekhar limit \citep{Hillebrandt2000,mazzali2007,livio_2018_progenitors_of_Ia}. A relation between the peak luminosity with the width of the light curve \citep{Phillips1993} makes SNe\,Ia standardisable candles and has led to their practical use as distance indicators and for the discovery of dark energy \citep{Riess1998,Perlmutter1998}.

The rise of the SN\,Ia light curve is caused by the deposition of the gamma-rays and positrons emitted in the decay of the \Nifs\ synthesised during the explosion \citep{Arnett1982,Kuchner1994,Mazzali_1998,mazzali2001}. The optical
photons created in this process remain trapped until the ejecta become optically
thin as they expand \citep{mazzali2001}, allowing their diffusion. Therefore, the peak of the light curve is directly proportional to the mass of \Nifs\ synthesised, while its width is related to the photon diffusion time, which is a function of ejected mass, kinetic energy, the radial distribution of \Nifs, and of the effective opacity, which is itself a function of temperature, density, and composition \citep{woosley}.

Even though the majority of SNe\,Ia constitute a nearly equivalent group of intrinsically bright events and their spectroscopic features are fairly similar, observations indicate a scatter in their spectroscopic properties \citep{Branch2001,Silvermanetal2012,Siebert2019,Jha2019}; see \citet{filippenko_1997_optical_spectra_of_Sn} for a review.

A question that arises is how distinct these events are. A clear separation could mean that they are of intrinsically different nature, while a continuity of properties would suggest quasisimilar events, with the observed diversity being caused by smoothly changing parameters. Important factors are the physical mechanism through which the white dwarf reaches ignition densities, the mass at explosion, and the explosion mechanism. 

Different regimes under which the burning flame propagates lead to different nucleosynthetic yields, different composition structures, and therefore different spectral features. Simulations of pure deflagration models were unable to reproduce \Nifs\ masses of $\sim 0.5\,\mathrm{M}_{\odot}$ and kinetic energies $\sim 10^{51}$\,ergs, as derived from observations \citep{mazzali2007}. In contrast, a pure detonation, in which the burning front propagates supersonically and ignites the fuel by compressive heating, incinerates the whole star to iron-group nuclei and cannot explain the presence of intermediate-mass elements (IMEs) in the ejecta outer layers. Alternative successful models have been proposed in which the deflagration front transits to a detonation at some specific density (deflagration to detonation transition, or DDT) \citep{khokhlov91DD}. One-dimensional simulations of delayed-detonation models have proven successful in reproducing many of the observed spectral features of SNe\,Ia, in particular, the presence of a layer of IMEs, the product of partial burning of carbon and oxygen.  These models can also account for the energy budget of the most energetic events and for the observed photospheric velocities. However, the exact physics of how this transition occurs is still a subject of extensive research \citep{woosley2007_burning_and_det_in_dist_regimes}.

The early-time spectra of normal SNe\,Ia are characterised by lines of singly-ionized IMEs such as Mg, Si, S, Ca, and iron-group elements (hereafter, Fe-gp). As time progresses, Fe lines increase in strength until they dominate the appearance of the spectrum a few weeks after maximum light \citep{filippenko_1997_optical_spectra_of_Sn, parrent2014_review_of_Ia_SN_spectra}.  However, in some ``peculiar'' events characterised by high luminosity (SN\,1991T, and SNe of its subgroup; \citealt[e.g.,][]{filippenko19921991t}), singly-ionised IMEs only start to appear near maximum light, never reaching the same intensity as in normal SNe\,Ia. Their early-time spectra are instead dominated by doubly-ionised species such as \FeIII\ and \SiIII. The presence of these lines requires high temperatures in the outer ejecta.

\begin{figure}
\includegraphics[trim={0 0 0 0},clip,width=0.47\textwidth]{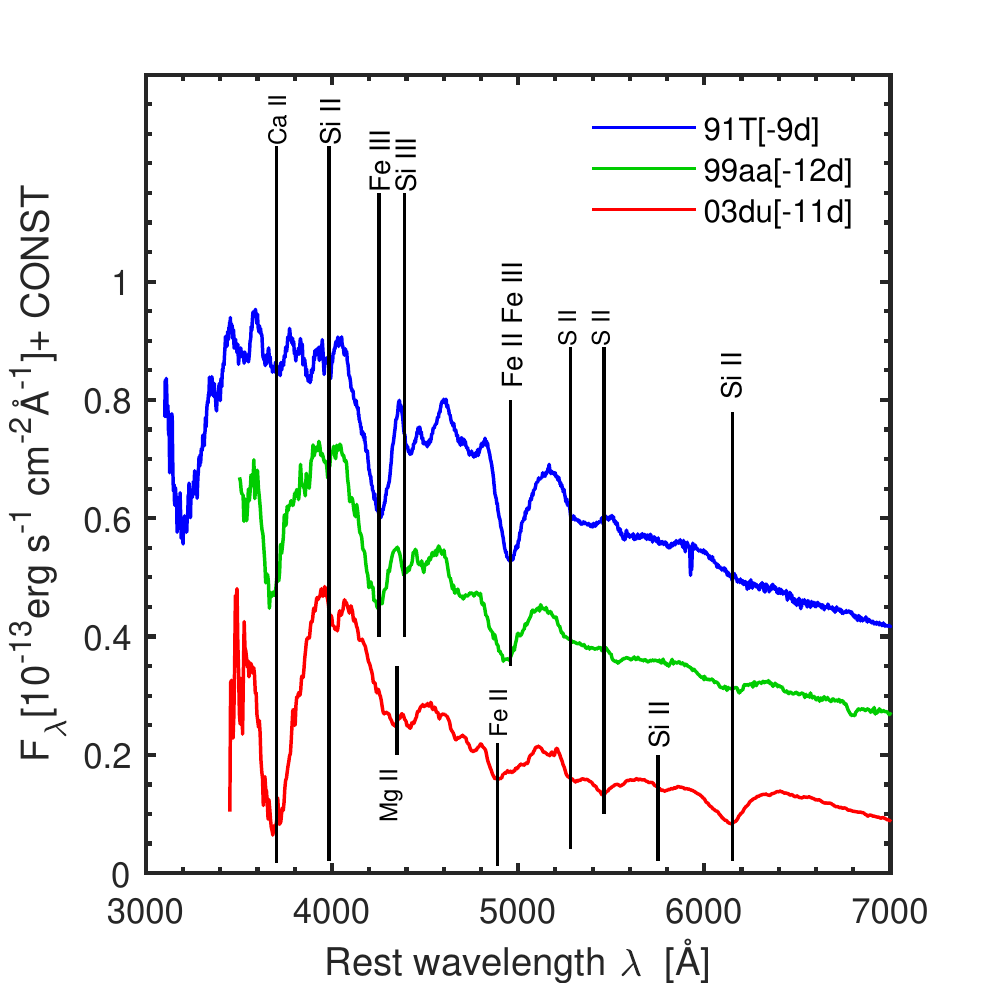}
 \caption{Early-time spectra of SN\,1991T (top), SN\,1999aa (middle), and the normal SN\,2003du (bottom). Major absorption features are marked. Epochs are given in days relative to $B$-band maximum brightness.}  
\label{fig1}
\end{figure}

\begin{figure}
\includegraphics[trim={0 0 0 0},clip,width=0.47\textwidth]{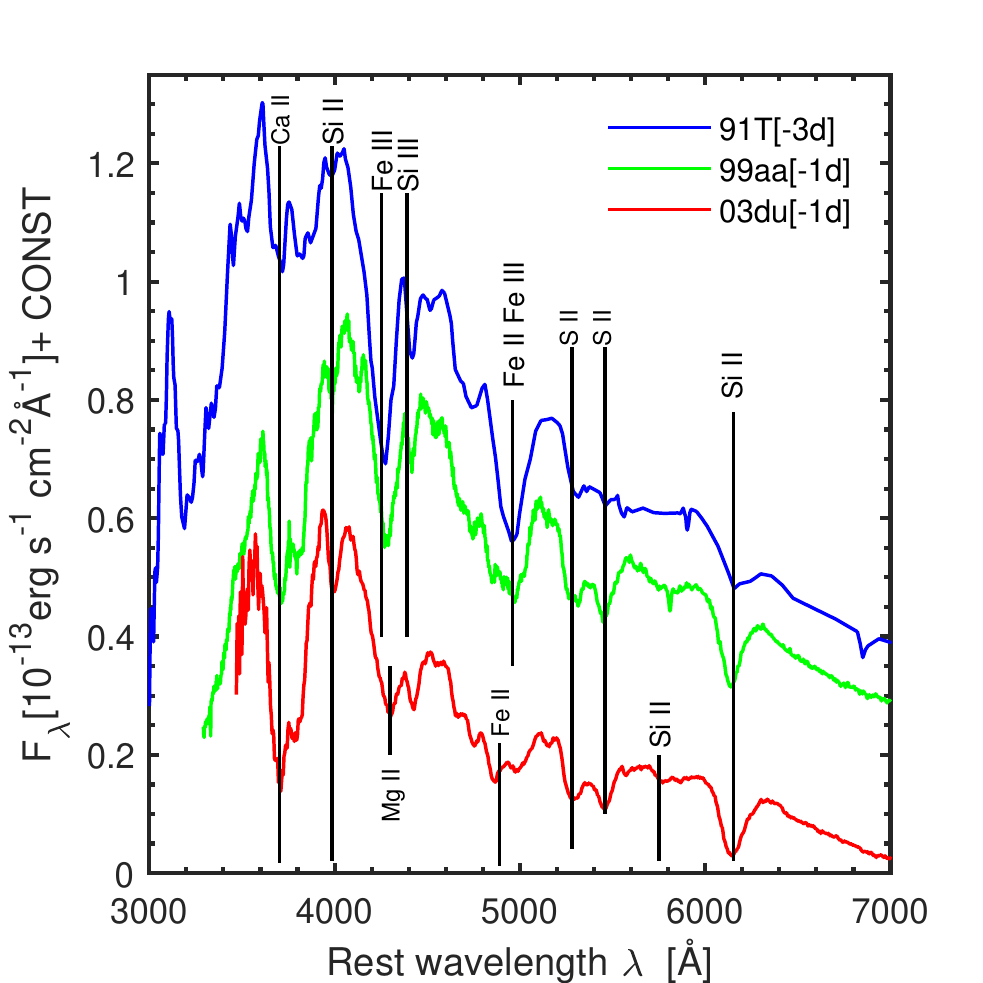}

\caption{Same as Fig. \ref{fig1}, but for near-maximum-brightness spectra.}  
\label{fig2}
\end{figure}

SNe\,Ia  with properties intermediate between those of SN\,1991T and normal SNe\,Ia have been discovered. One case in particular is that of SN\,1999aa, the subject of this study \citep{garavini2004,Jhaetal2006,Matheson2008}. Similar to SN\,1991T, SN\,1999aa was a slow decliner, with $\Delta m_{\mathrm{15}}(B)$ measurements ranging from 0.75\,mag \citep{Krisciunas_2000_photom} to 0.85\,mag \citep{Jhaetal2006}. The earliest spectra of SN\,1999aa resemble those of SN\,1991T in being dominated by \FeIII\ lines and by the weakness of singly-ionised IME lines, in particular \SiII\,6355\,\AA.  However, unlike SN\,1991T, they showed a high-velocity \CaII\,H\&K feature. SN\,1999aa morphed to looking like a normal SN\,Ia earlier than did SN\,1991T. In fact, one week before $B$ maximum, \SII\,5468, 5654\,\AA\ and \SiII\,6355\,\AA\ were already visible in SN\,1999aa. Figures \ref{fig1} and \ref{fig2} show optical spectra of SN\,1999aa compared to SN\,1991T and the spectroscopically normal SN\,2003du, respectively $\sim 10$\,days before and near $B$ maximum.   

It has been suggested that SN\,1999aa and similar SNe\,Ia (e.g., SNe\,1998es, \citealt{Ayani1998}; 2012cg, \citealt{Silverman2012cg}; 1999dq, 2001eh, 2002hu, 2006cz, and 2008Z, \citealt{SilvermanII2012}) constitute a subclass of their own. \citet{SilvermanII2012} estimate the rate of SN\,1999aa-like SNe\,Ia to be comparable to that of SN\,1991T-like events. 

A theoretical understanding of SN\,1999aa should help clarify the spectroscopic sequence from normal to SN\,1991T-like events.  A first step toward this is to derive the composition and stratification of the ejecta. This can be done using the so-called ``abundance tomography'' technique \citep{Stehle2005}, which involves modelling a temporal series of spectra to reproduce their features consistently. At early times, the spectra are characterised by a pseudocontinuum on which P~Cygni profiles of the lines that dominate near the momentary photosphere are superimposed. As the ejecta expand, the photosphere recedes inward and reveals progressively deeper layers. This approach was successfully used to model several SNe\,Ia: SN\,2002bo \citep{Stehle2005}, SN\,2004eo \citep{Mazzali2008}, SN\,2003du \citep{Tanaka2011}, SN\,1991T \citep{Sasdelli2014}, and SN\,1986G \citep{Ashall2016}.

Here, we use the abundance tomography technique to investigate the properties of SN\,1999aa. In Section \ref{sec2} we describe the data used, and in Section \ref{sec3} we explain the modelling methods. We present our modelling results in Sections \ref{sec4} and \ref{sec5}. In Section \ref{sec6} we discuss the abundance tomography results. We use the derived abundances to compute a bolometric light curve in Section \ref{sec7}, and Section \ref{sec8} discusses our results. Our conclusions are drawn in Section \ref{sec9}.

\section{DATA}
\label{sec2}

SN\,1999aa was discovered independently by \citet{Nakano1999}, \citet{Armstrong1999}, and \citet{Qiao1999}. The host galaxy is NGC\,2595, a barred spiral of morphological classification SAB(rs)c with redshift $z = 0.0144$ \citep{ngc2595redsfift_Epinat,vandriel2016}. Distance moduli to the galaxy based on the Tully-Fisher relation range from \cha{$\mu=32.30 \pm 0.53$\,mag} \citep{distNGC2595botinelli32.3} to $\mu=34.44 \pm 0.47$\,mag \citep{distNGC2595Theureau34.07}. Distances using the light curve of SN\,1999aa vary from $\mu=33.43 \pm 0.16$\,mag \citep{ngc2595classamanullah} to $\mu=34.58 \pm 0.24$\,mag \citep{distngc2595LC_Riess34.58}.

Photometric data are taken from \citet{Jhaetal2006}, \citet{Krisciunas_2000_photom}, \citet{Qiao1999}, \citet{Armstrong1999}, \citet{Yoshida_1999aa}, and \citet{Altavilla_2004}. Late-time unpublished data are based on observations collected with the Optical Imager Galileo (OIG) at Telescopio Nazionale Galileo (TNG) -- La Palma. The TNG + OIG $UBVRI$ frames were reduced following standard procedures and made use of the \textsc{ecsnoopy} package \citep{snoopy} using the point spread function (PSF) fitting technique for the SN measurement. The $BVRI$ SN magnitudes were then calibrated with reference to the magnitudes of field stars retrieved from \citet{Krisciunas_2000_photom}, while for the $U$ band, we converted the SDSS catalog magnitudes of the local sequence into Johnson $U$ following \citet{Chonis_2008_SDSS_Zeropoints}. The final TNG + OIG magnitudes are shown in Table \ref{tab1}, where the mean photometric errors, estimated with artificial-star experiments, are given in parentheses.

The spectra used in this study are available at the Weizmann Interactive Supernova Data Repository (WISeREP) \citep{wiserep}; they are listed in Table \ref{tab2}. The spectra were calibrated against photometric observations. Calibration was performed in the $U$, $B$, $V$, and $R$ bands by multiplying the spectra with a line of constant gradient or with a low-order smoothed spline. We ensured that the flux in the spectra in any passband did not vary by more than $\sim 10$\% from the observed flux in that filter passband. 
\vspace{-10pt}

\begin{table*}
\centering 
\caption{Late-time photometry of SN\,1999aa.}
\label{tab1} 

\begin{threeparttable}

\begin{tabular}{lllcccccc}

  
\hline
   UT Date   &  JD$^a$ & Epoch$^b$ & $U$ ($1\sigma$) & $B$ ($1\sigma$) & $V$ ($1\sigma$) & $R$ ($1\sigma$) & $I$ ($1\sigma$)  \\
          &          &   (days) &    (mag) & (mag) & (mag) & (mag) & (mag)    \\
  \hline
 
 03/11/1999$^c$ & 485.170 & $+$248.9 & 22.202 (0.086) & 20.987 (0.075) & 20.811 (0.045) & 22.019 (0.127) & ...  \\
 11/12/1999$^c$ & 523.170 & $+$286.4  & ... & 21.381 (0.044) & 21.192  (0.029) & 22.705 (0.079) & ...   \\
 14/12/1999$^c$ & 526.098 & $+$289.2  & 23.067 (0.067) & 21.486 ( 0.025) & 21.189 (0.026) & 22.769 (0.099) & 21.494 (0.071)  \\
 05/01/2000$^c$ & 548.086 & $+$310.9  & 23.414 (0.079) & 21.844 (0.028)  & 21.522 ( 0.027) & 23.059 ( 0.050) & 22.112 (0.094)  \\
 
\hline
\end{tabular}
$^a$ JD$-$2,451,000; 
$^b$ Rest-frame time, since $B$ maximum; 
$^c$ Instrument: TNG + OIG
\end{threeparttable}
\end{table*}

\begin{table}
\setlength{\tabcolsep}{2pt}
\caption{Spectra of SN\,1999aa and modelling parameters.}
\label{tab2} 


\scalebox{0.95}{
\hskip-0.5cm\begin{tabular}{llcccccc}

\hline
   UT Date   &  JD$^a$ & Epoch$^b$ & Telescope/Instr. & log\,$L$ & $v$ & $T$(rad) \\
    1999      &          &   (days)         & &[\Lsun] & (\kms) & (K) \\
  \hline
 12/02$^c$ & 221.5 & $-$11.0  & Lick 3\,m / Kast & 9.220 & 12600 & 12310 \\
 13/02$^d$ & 222.5 & $-$10.0  & NOT 2.6\,m / ALFOSC & 9.315 & 12300 & 12735 \\
 14/02$^e$ & 223.83 & $-$8.8  & FLW 1.5\,m / FAST  & 9.440 & 12000 & 12946 \\
 15/02$^e$ & 224.78 & $-$7.8 & FLW 1.5\,m / FAST  & 9.470 & 11500 & 13335 \\
 16/02$^e$ & 225.74 & $-$6.9 & FLW 1.5\,m / FAST  & 9.510 & 11200 & 13440 \\
 17/02$^d$ & 226.5  & $-$6.1  & APO 3.5\,m / DIS   & 9.530 & 10950 & 13378 \\
 18/02$^e$ & 227.75 & $-$4.9  & FLW 1.5\,m / FAST  & 9.590 & 10550 & 13695 \\
 19/02$^e$ & 228.78 & $-$3.8 &  FLW 1.5\,m / FAST  & 9.610 & 10300 & 13490 \\
 20/02$^e$ & 229.79 & $-$2.9  & FLW 1.5\,m / FAST  & 9.610 & 10000 & 13239 \\
 21/02$^e$ & 230.79 & $-$1.9 & FLW 1.5\,m / FAST  & 9.595 & 9600  & 13029 \\
 23/02$^c$ & 232.5  & $-$0.2 & Lick 3\,m  / Kast  & 9.590 & 9300  & 12296 \\
 25/02$^d$ & 234.5  & $+$1.8 & Lick 3\,m  / Kast  & 9.560 & 8900  & 11380 \\
 03/03$^d$ & 240.5  & $+$7.7 & MDM 2.4\,m / Mk III & 9.310 & 7150 & 12778 \\
 09/03$^c$ & 246.5  & $+$13.6 & Keck-1 / LRIS    & 9.170 & 5500 & 11663 \\
 12/03$^c$ & 249.5  & $+$16.6 & Lick 3\,m / Kast  & 9.130 & 4250 & 11162  \\
\hline
\end{tabular}}

\small
$^a$ JD$-$2,451,000 \\
$^b$ Rest-frame time, since $B$ maximum \\
$^c$ \cite{Silvermanetal2012}\\
$^d$ \cite{garavini2004}\\
$^e$ \cite{Matheson2008}\\
\end{table}

\section{MODELLING TECHNIQUES}
\label{sec3}

\begin{figure}
\hspace*{-0.6cm}
\includegraphics[trim={0 0 0 0},clip,width=0.54\textwidth]{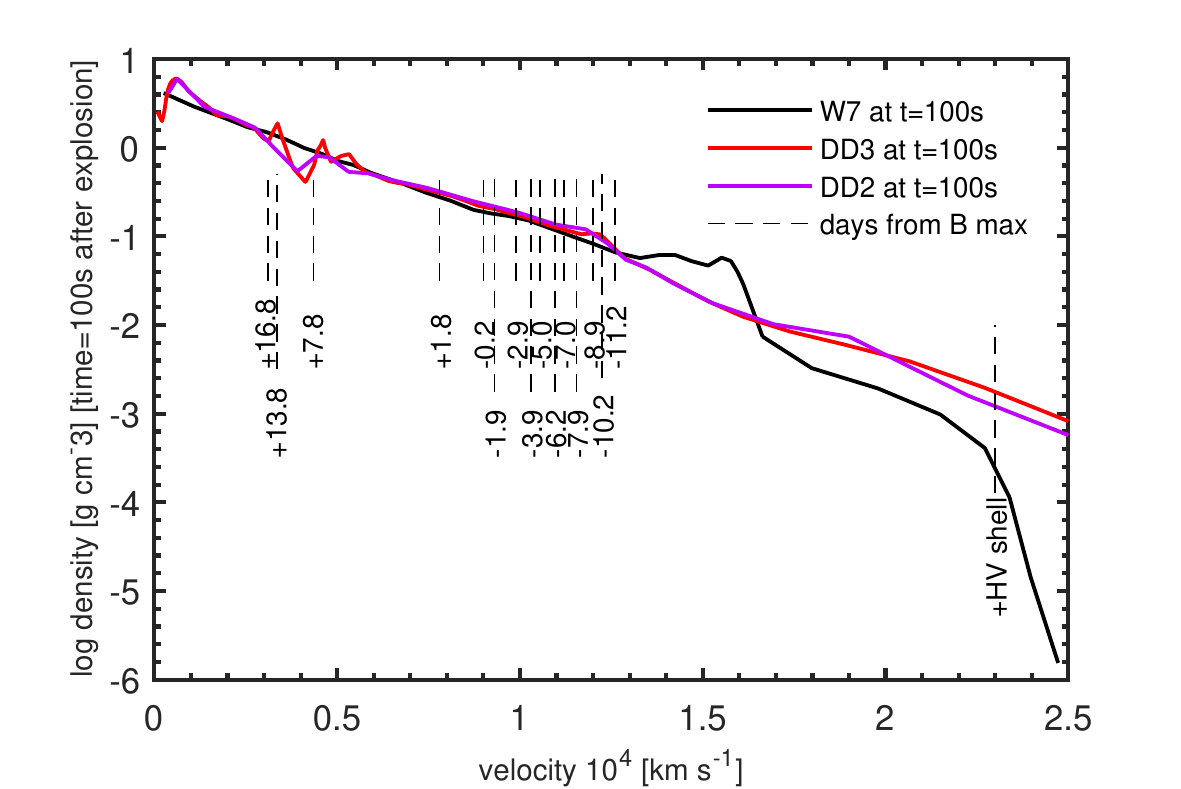}
\caption{The three density profiles used in the modelling: W7 \citep{nomoto1984}, DD2, and DD3 \citep{iwamoto1999}. Vertical dashed lines mark the photospheres of the synthetic spectra.}  
\label{fig3}
\end{figure}

Spectra in the photospheric phase have been modelled using a Monte Carlo spectrum synthesis code \citep{Mazzali1993,Lucy1999a,Lucy1999}. 
The code assumes a sharp photosphere. As the ejecta expand homologously, the photosphere recedes in velocity space and consequently in mass coordinate. Thermal equilibrium is assumed. Photons emitted from the photosphere propagate through the expanding ejecta and interact with the gas through line absorption, including line branching \citep{Mazzali2000}, or electron scattering. The required input parameters are the density structure of the SN ejecta, the emergent bolometric luminosity $L_{\mathrm{Bol}}$, the photospheric velocity $v_{\mathrm{ph}}$, the time from the explosion $t_\mathrm{0}$, and the abundances of the elements as a function of depth above the photosphere.

The distance and the extinction to the SN are needed in order to scale the flux. Since the distance to NGC\,2595 is not known accurately, we treat it as a free parameter, within the range allowed by the literature. We tested several values of the distance modulus ($\mu$) for three different spectra. For larger distances the high luminosity causes a high temperature, which in turn leads to unrealistic ionisation. The opposite happens for distances that are too small. The best models are obtained with $\mu = 34.00$\,mag, \cha{which is very close to the mean value calculated using the various distance moduli reported in the literature, $\bar{\mu}$=33.975$\pm 0.34\, $ mag}. We adopt an extinction value $E(B-V)=0.04$\,mag \citep{Schlegel_milkywayextinction_1998} for the Milky Way and assume $E(B-V)=0.00$\,mag for the host galaxy \citep{Krisciunas_2000_photom}. We also tested different rise times, between 19 and 21\,days; the best results are obtained with a value of 20\,days.

We use three different density-velocity distributions: the classical fast deflagration model, W7 \citep{nomoto1984}, and two more-energetic delayed-detonation models, DD2 and DD3 from \citet{iwamoto1999}. These density profiles are shown in Fig. \ref{fig3}. 

Having fixed $\mu$, $t_\mathrm{0}$, and $E(B-V)$, the modelling starts with the earliest spectrum. Different values of $L_{\mathrm{Bol}}$ are tried until the synthetic spectrum matches the observed one in flux. After that, $v_{\mathrm{ph}}$ is iterated to match the position of the spectral features and the overall temperature. In parallel, the abundances are modified until the model matches the observation.

For the following spectrum in the sequence, a new, smaller $v_{\mathrm{ph}}$ is defined. This will introduce a new shell where new abundances can be determined. This process is repeated for each spectrum. As the spectra evolve, deeper layers are revealed and the abundance stratification is gradually built.

\section{THE PHOTOSPHERIC PHASE}
\label{sec4}

\begin{figure*} 
\includegraphics[trim={0 0 0 0},clip,width=0.86\textwidth]{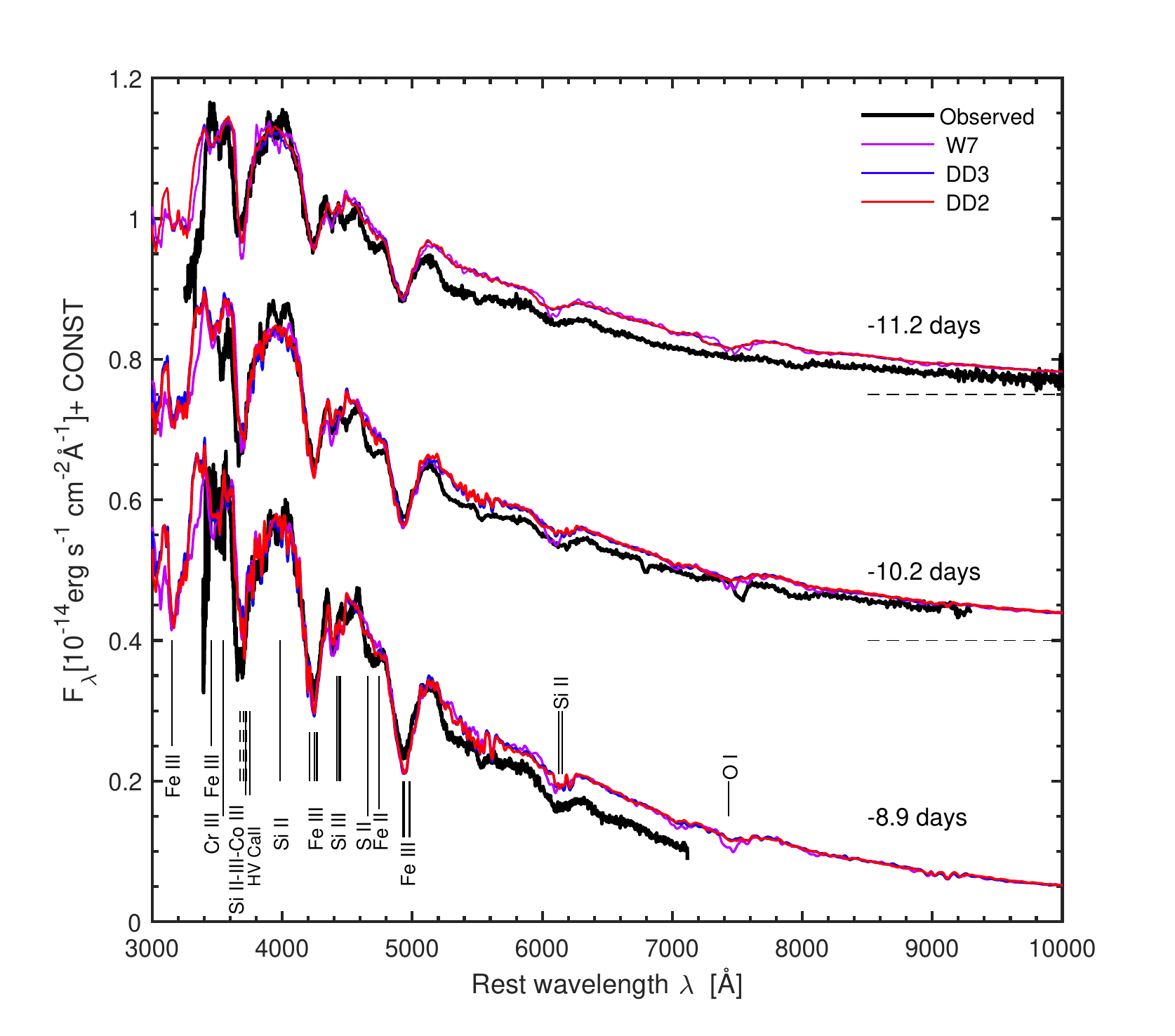}
\caption{Early-time spectra. Epochs are shown with reference to $B$ maximum. Features in the $-10.2$\,day spectrum near 6780\,\AA\ (very weak) and 7500\,\AA\ (weak) are telluric. The spectra have been shifted in flux by a constant value. The horizontal dashed line marks the zero flux for each epoch.}  
\label{fig4}
\end{figure*}

We modelled 15 spectra, from day $-11$ to day $+14$ from $B$ maximum.  The input parameters are shown in Table \ref{tab2}. The synthetic spectra corresponding to the three explosion models we use are shown in Figs. \ref{fig4}, \ref{fig8}, \ref{fig9}, \ref{fig10}, and \ref{fig11}, overlaid on the observed spectra.

\begin{figure}
\includegraphics[trim={0 0 0 0},clip,width=0.47\textwidth]{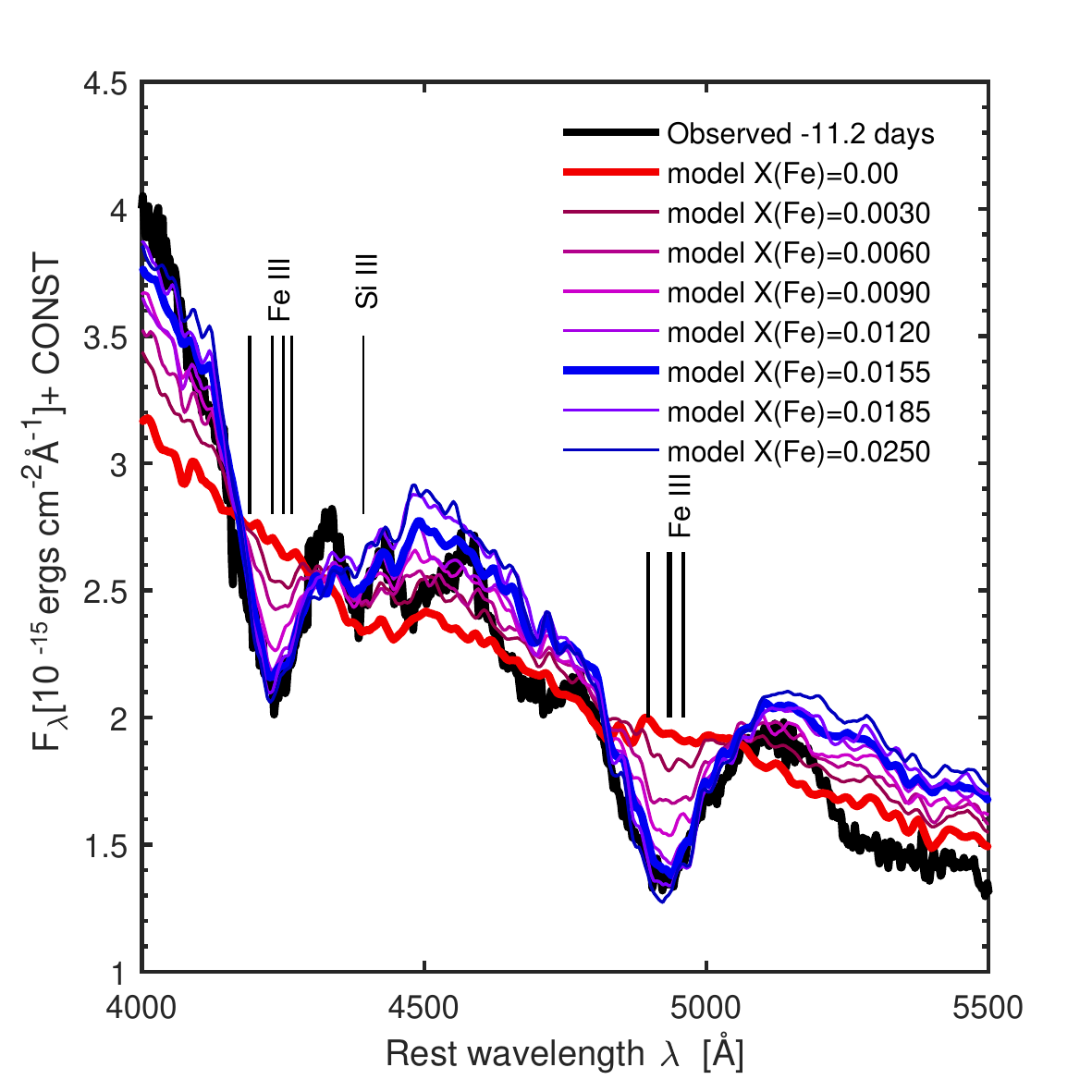}
\caption{A close-up view of the day $-11$ spectrum showing the region where the strongest \FeIII\ lines are seen. Adding small packets of stable Fe, the Fe mass fraction can be inferred. }  
\label{fig5}
\end{figure}

\subsection{The early-time spectra}
\label{sec4.1}

In Fig. \ref{fig4} we show models for the earliest spectra, ranging from ${-}$11 to ${-}$9 days from $B$ maximum. The synthetic spectra reproduce the observed features well. In particular, they exhibit deep absorption lines of \FeIII\ and \SiIII, and the overall flux matches the observed flux, except in the redder bands. The velocity of the photosphere starts at 12,600\,\kms\ and decreases to 12,000\,\kms.

\textit{Fe-group elements:} A small amount of iron is needed at the outer shells to reproduce the deep observed features near 4250 and 5000\,\AA. This is stable Fe; at these early stages, \ce{^{56}Ni} would not have had time to decay significantly into \ce{^{56}Fe}.  The mass fraction of Fe at $v > 12,600$\,\kms needed to reproduce the observed features is $\sim 0.015$--0.018 (Fig.\ref{fig5}). The presence of stable Fe in the outer shells has been reported in other SNe\,Ia. \citet{Sasdelli2014} give a solar abundance \citep[$X$(Fe$_\odot$) = 0.001][]{asplund-2009-solar-abundance} for SN\,1991T, while \citet{Tanaka2011} report values of $\sim 0.003$--0.005 for SN\,2003du. Our values are significantly supersolar.

Small amounts of \ce{^{56}Ni}, Ti, and Cr are needed at these epochs to block the ultraviolet (UV) flux and redistribute it redward. The abundance of \ce{^{56}Ni} is not constrained at these epochs, as no visible line in the spectrum is reproduced by \ce{^{56}Ni} or Co alone. Unfortunately, the spectra of SN\,1999aa do not extend bluer than $\sim 3400$\,\AA, where a prominent feature dominated by \CoIII\ should be expected at $\sim 3200$\,\AA\ \citep{mazzali1995,Stehle2005,Sasdelli2014}.

\textit{Calcium:} The early-time spectra of SN\,1999aa show a deep absorption line near 3700\,\AA, with strength intermediate between that of normal SNe\,Ia and SN\,1991T \citep{garavini2004}. This feature is due to high-velocity \CaII\,H\&K \citep{mazzali2005HVfeatures}. 

We are able to produce it with $X$(Ca) $\approx 0.0035$ at $v > 21,000$\,\kms\ with the W7 model. The DD2 and DD3 models have more mass at high velocity, and therefore $X$(Ca) $\approx 0.00025$ is sufficient \citep[see][]{Tanaka2011}. This is much less than the Ca abundance reported in some spectroscopically normal SNe\,Ia \citep{tanaka_outermostejecta_2008}. However, those SNe exhibit a much stronger \CaII\,H\&K lines and a much earlier appearance of the \CaII\ NIR feature than in both SN\,1999aa and SN\,1991T. On the other hand, the abundance we obtained is similar to that obtained for SN\,1991T, for which \citet{Sasdelli2014} estimate a Ca abundance $< 0.0003$ at $v > 17000$\,\kms\, using the DD3 density profile. 
 Regardless of the density profile used, our results suggest that the abundance of Ca at high velocities is supersolar \citep[$X$(Ca$_{\odot}$) = 0.00006, ][]{asplund-2009-solar-abundance}.

\begin{figure*}
\includegraphics[trim={0 0 0 0},clip,width=0.88\textwidth]{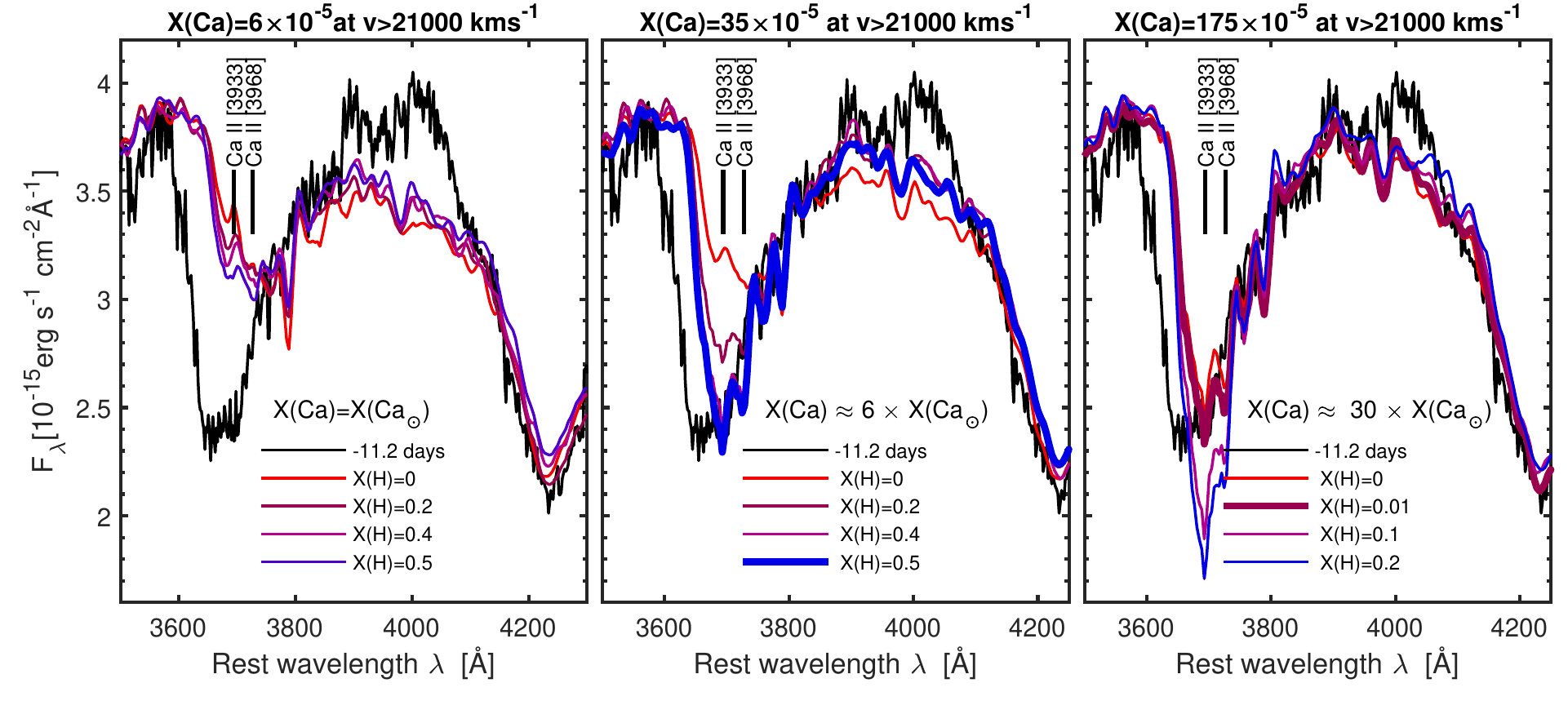}
\caption{A close-up view of the \CaII\,H\&K region, showing the observed spectrum at day $-$11 compared to models computed with the W7 density profile using different abundances of Ca coupled with various amounts of H in the outermost shells. }  
\label{fig6}
\end{figure*}

The strength of the \CaII\,H\&K feature is very sensitive not only to the Ca abundance at high velocity, but also to the parameters that directly affect ionisation, in particular the electron density. The presence of free electrons decreases the ionisation and favours singly-ionised species \citep{mazzali2005_HVF_1999ee_H_electrondensity,mazzali2005HVfeatures}. Adding H results in a higher electron density. Following \citet{Tanaka2011}, in Fig. \ref{fig6} we show how the \CaII\,H\&K feature can be reproduced with different Ca abundances coupled with different amounts of H at the outermost shells ($v > 21,000$\,\kms). However, because of the degeneracy between the Ca abundance and the electron density, it is not possible to determine the Ca mass fraction. \cha{ Hydrogen may result from the interaction of the ejecta with the circumstellar medium \citep{mazzali2005HVfeatures} or may be a remnant  of accretion on the surface of the WD \citep{livio_2018_progenitors_of_Ia} . Even though small amounts of H are sufficient to reduce the ionization at the dilute outermost layers , and therefore create the HVFs ubiquitously observed in SNe\,Ia spectra \citep{mazzali2005HVfeatures},  larger amounts ($X(H)\gtrapprox$~0.3), will give rise to an H$\alpha$} \cha{feature that is not seen in the observed spectrum. The lack of H signatures can be taken as an argument against the single degenerate scenario \citep{marietta_Ia_impact_of_sec_star, PANAGIA-lack_of_H_frm_radio} but it is not enough to rule it out. \citep {justham_SD_wo_H,hachisu_SDprogenitor}\citep[for a review, see][]{livio_2018_progenitors_of_Ia}}.

\textit{Silicon, Sulphur, Magnesium:} The \SiII\ 6355\,\AA\ line is much weaker in the earliest spectra of SN\,1999aa than in normal SNe\,Ia. It grows in strength as the spectra evolve. However, the feature near 4400\,\AA, which is due to \SiIII\ 4553, 4568, 4575\,\AA, is prominent in the earliest spectra, as the high temperature favours doubly-ionised species. These Si lines are well reproduced in the synthetic spectra. The Si mass fraction is 0.025 at $v > 12,600$\,\kms, but it rapidly increases to 0.73 at $v > 12,300$\,\kms.

The two \SII\ features at 5468 and 5654\,\AA\ are not present at these early epochs, and only start to show at day $\sim -6$. 
The \MgII\,4481\,\AA\ line is never visible in the spectra of SN\,1999aa, as that region is dominated by \FeIII\ lines. 

\begin{figure}
\includegraphics[trim={0 0 0 0},clip,width=0.47\textwidth]{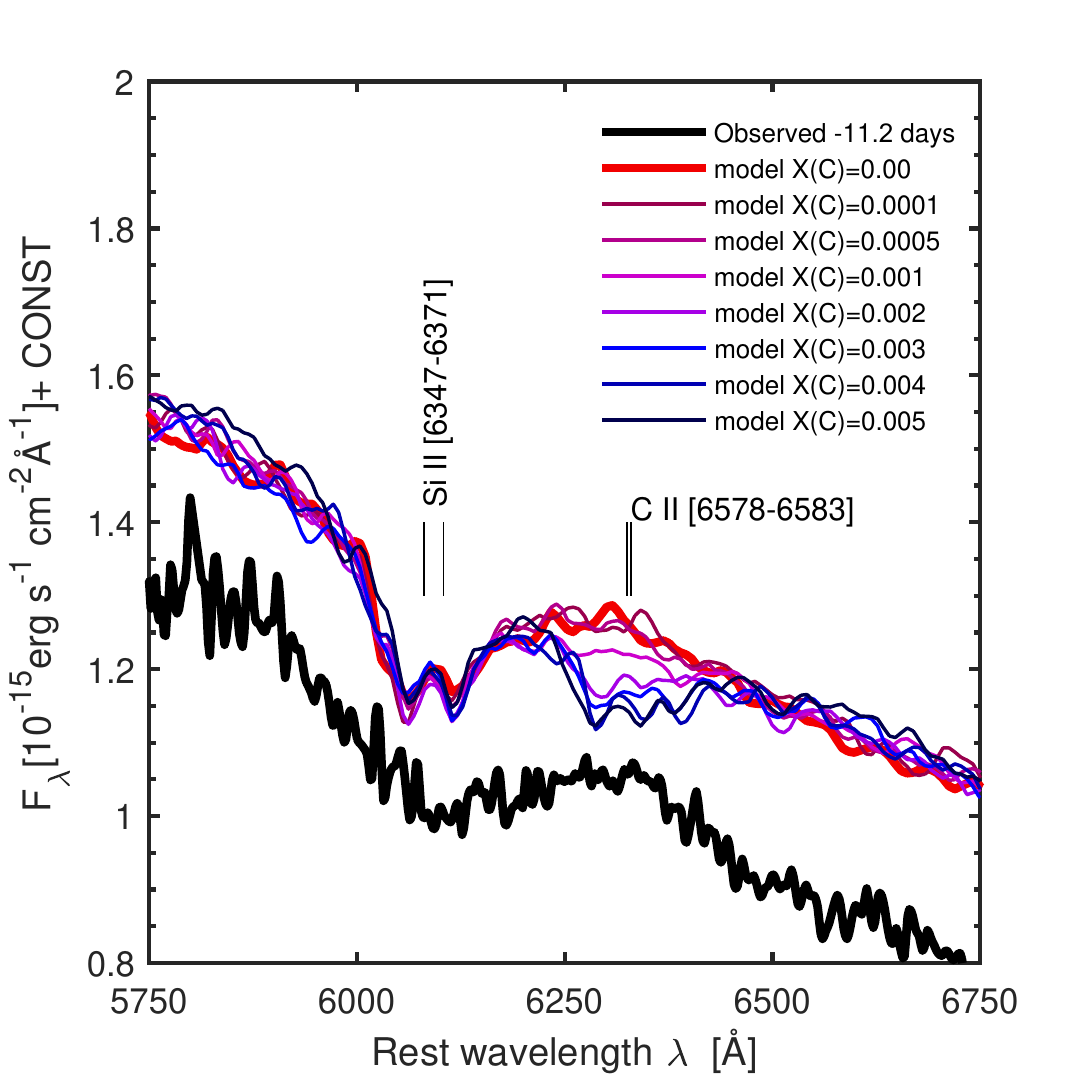}\\
\caption{A close-up view of the spectrum showing the \SiII--\CII\ region. Small amounts of carbon can create an absorption line redward of the \SiII\ line. A prominent C line is not seen in the observed spectrum. }  
\label{fig7}
\end{figure}

\begin{figure*}
\includegraphics[trim={0 0 0 0},clip,width=0.86\textwidth]{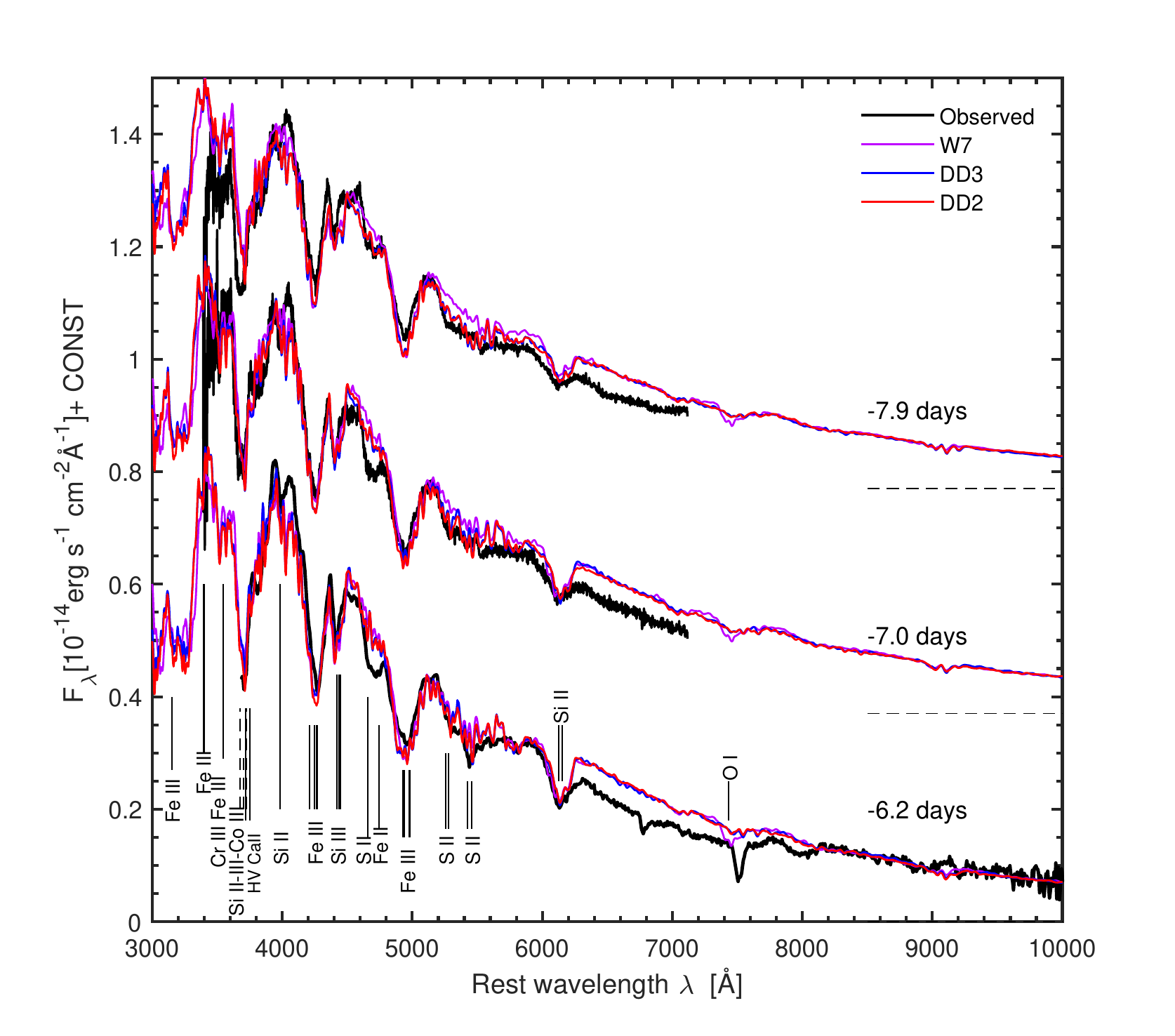}\\
\caption{Spectra one week before $B$-band maximum. Epochs are shown with reference to $B$ maximum. Features in the $-6.2$\,day spectrum near 6780\,\AA\ (weak) and 7500\,\AA\ are telluric. The spectra have been shifted in flux by a constant value. The horizontal dashed line marks the zero flux for each epoch.}  
\label{fig8}
\end{figure*}

\begin{figure*}
\includegraphics[trim={0 0 0 0},clip,width=0.86\textwidth]{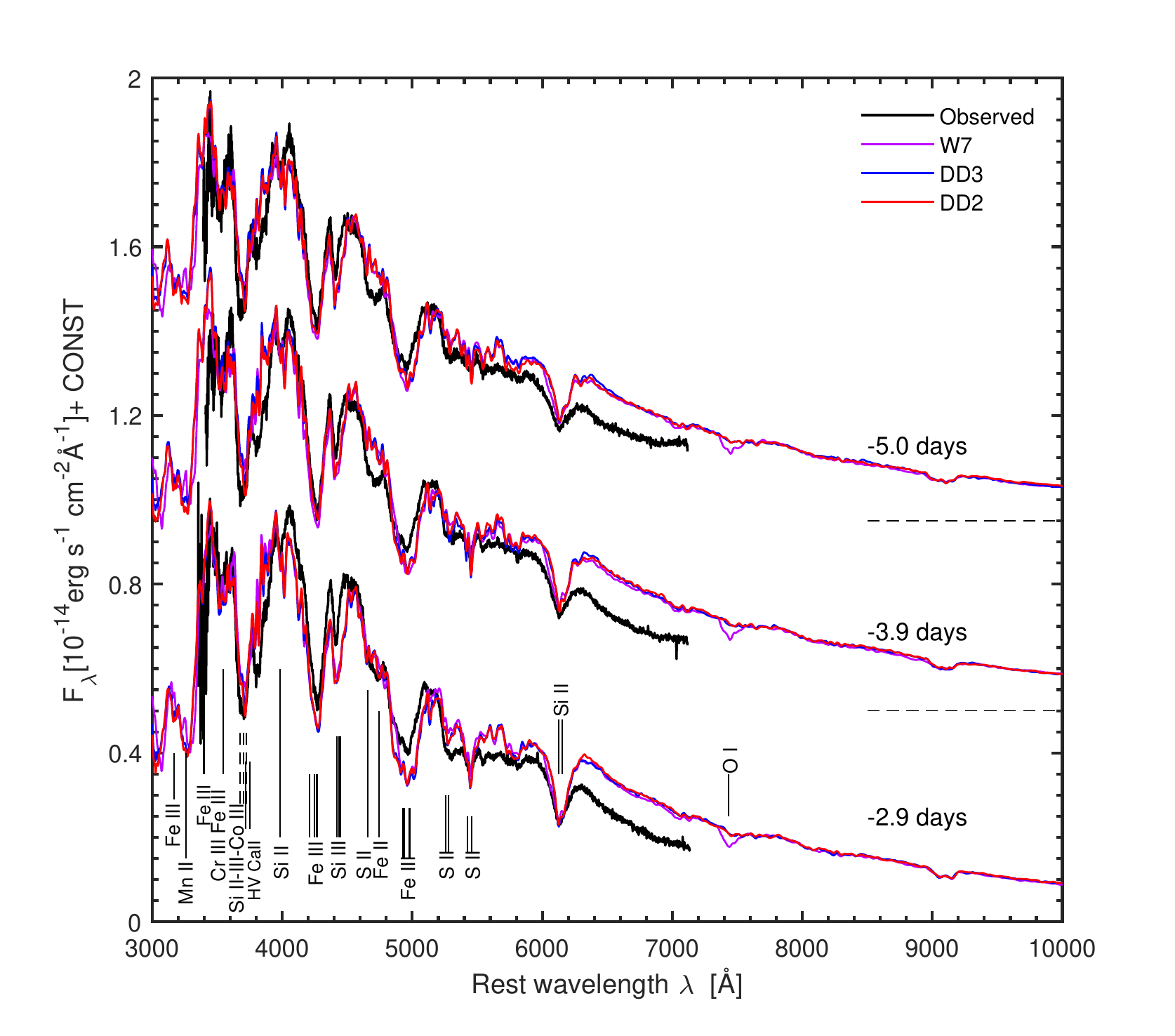}\\
\caption{Pre-maximum-brightness spectra. Epochs are shown with reference to $B$ maximum. The spectra have been shifted in flux by a constant value. The horizontal dashed line marks the zero flux for each epoch.}  
\label{fig9}
\end{figure*}

\textit{Carbon, Oxygen:} A \CII\, 6578, 6583\,\AA\ line has been detected in some SNe\,Ia \citep{Mazzali2001_C_and_Si, Parrentcarbonfeatures2011}. It can be observed on top of the \SiII\ 6355~\AA\ P~Cygni emission, but it is not a common feature. This line is not visible in SN\,1999aa. An upper limit to the carbon abundance of $\sim 0.0005$ by mass at $v > 12,600$\,\kms\ can be determined (Fig.~\ref{fig7}). The absence of C in the outermost layers favours delayed-detonations models \citep{khokhlov91DD,marioncarbon}. \cha{It is also possible that carbon is present, but most of it is in a doubly ionized state and therefore does not produce a visible feature in the spectrum. It is difficult to excite any lower level of any optical line of \CIII\ at the temperatures of even a luminous SN Ia. Carbon may also be present at much higher velocities, but in order to investigate this, we need earlier observations which unfortunately are not available for SN\,1999aa.}

The synthetic spectra using W7 show a shallow \OI\ 7744\,\AA\ line (see Fig. \ref{fig4}). This feature is not seen in spectra of SN\,1999aa. The DD2 and DD3 profiles have less mass in the region between 13,000 and 16,000\,\kms, where the line is formed (see Fig. \ref{fig3}), and produce a much shallower feature that matches better the observed spectra (days $-$11 and $-$10). The oxygen abundance can be constrained by considering the abundances of other elements that are present at high velocity (Si, S, Ca, Fe, and \ce{^{56}Ni}). Oxygen dominates at $v > 12,600$\,\kms, but is absent already at $v \approx 12,300$\,\kms.  This behaviour is remarkably similar to that of SN\,1991T \citep{Sasdelli2014}. In contrast, \citet{Tanaka2011} report the presence of oxygen down to a velocity of 9400\,\kms for the normal SN\,Ia 2003du.

\subsection{Pre-maximum-light spectra}

Figs. \ref{fig8} and \ref{fig9} show spectra ranging from $-$8 to $-$3 days from $B$ maximum. The photospheric velocity evolves from 11,500 to 10,000\,\kms. 

\textit{Fe-group elements:} The \FeIII\ lines observed near 4300 and 5000\,\AA\ increase in strength. Our synthetic spectra reproduce their  evolution. At these epochs, the fraction of Fe originating from \Nifs\ decay becomes significant. At day $-$7, it already constitutes 20\% of the total Fe mass fraction, increasing to 30\% at day $-$3. We obtain good fits for $X$(Fe$_{\mathrm{stable}}) \approx 0.1$ in the shells that are probed.  
The abundance of \Nifs\ increases from 0.05 at $v=11,500$\,\kms\ to 0.53 at $v=10,950$\,\kms.

\textit{Calcium:} The synthetic spectra reproduce well the \CaII\,H\&K feature. The \CaII\ near-infrared (NIR) triplet is still not seen in the observed spectra at day $-$6, and this is confirmed in our synthetic spectra. The near-photospheric Ca abundance at these epochs is $\sim 0.0015$ for all models.

\textit{Silicon, Sulphur:} The \SiII\ 6355\,\AA\ line gets deeper with time. This is well replicated in our synthetic spectra, as are the \SiIII\ feature near 4400\,\AA\ and the \SiII\ 4130\,\AA\ line. The Si abundance is 0.72 at $v \approx 11,500$\,\kms, and it decreases to 0.26 at $v \approx 10,000$\,\kms.
The two \SII\ lines at 5468 and 5654\,\AA\ start to show at day $-$6 and grow stronger with time. The sulphur abundance is 0.15 at $v \approx 11,500$\,\kms, decreasing slightly to lower velocities.

\textit{Carbon, Oxygen:} C and O are not needed at these epochs. Any abundance of C would produce a line that is not seen in the observed spectra. Oxygen is not needed because the Fe-group elements and IMEs are sufficient to complete the composition at these velocities.

\begin{figure*}
\includegraphics[trim={0 0 0 0},clip,width=0.86\textwidth]{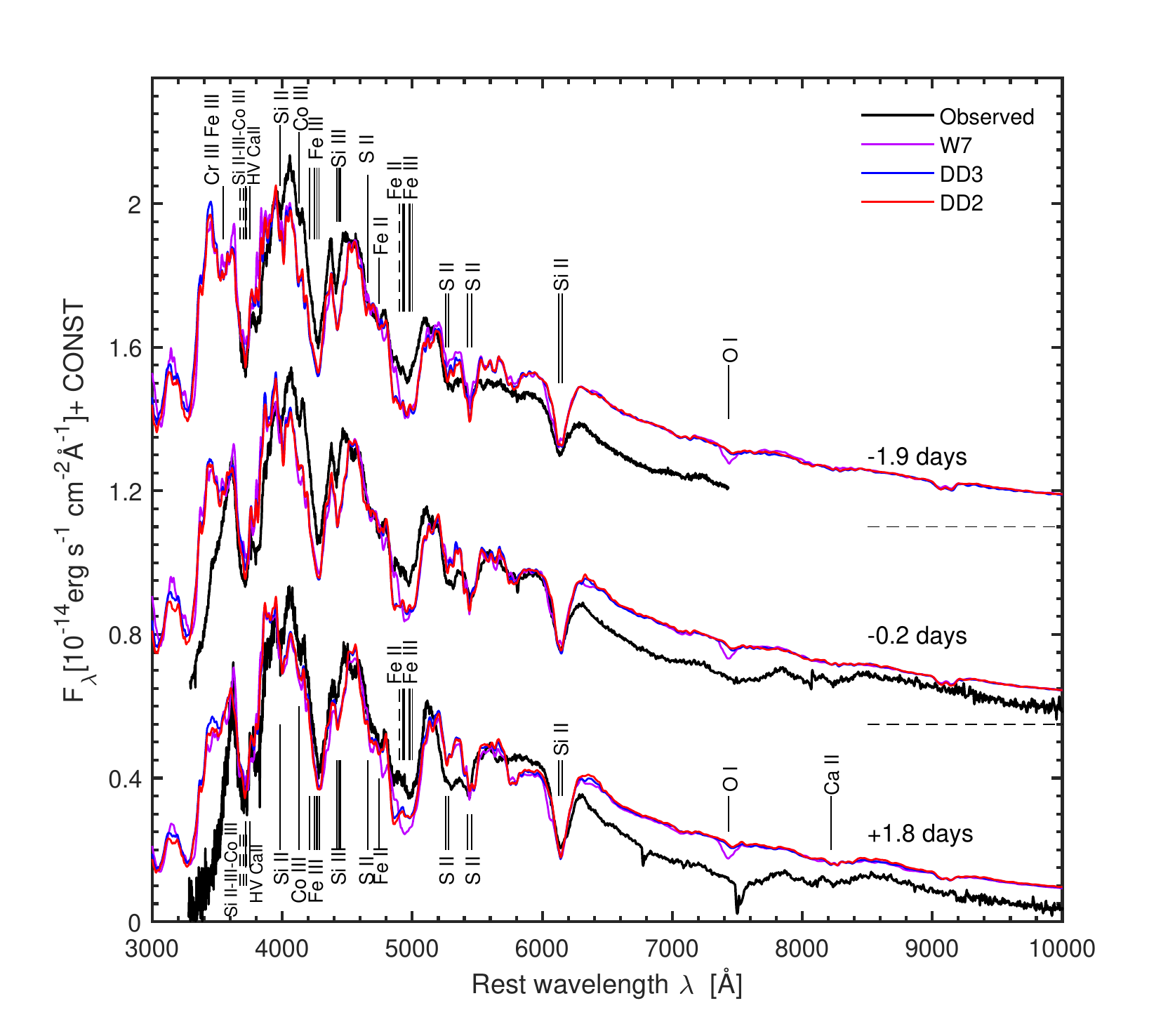}\\
\caption{Near-maximum-light spectra. Epochs are shown with reference to $B$ maximum. Features in the $+1.8$\,day spectrum near 6780\,\AA\ (weak) and 7500\,\AA\ are telluric. The spectra have been shifted in flux by a constant value. The horizontal dashed line marks the zero flux for each epoch.}  
\label{fig10}
\end{figure*}

\subsection{Spectra at maximum brightness}

Fig. \ref{fig10} shows spectra ranging from $-2$ to  $+2$ days from $B$ maximum. The photospheric velocity evolves from 9600 to 8900\,\kms. The synthetic spectra predict excess flux redward of $\sim 6000$\,\AA.  At these epochs, as the photosphere recedes inside the \Nifs-dominated shell, energy is partially deposited above the photosphere, and the assumption of blackbody emission at the photosphere is not entirely correct. 

\textit{Fe-group elements:} At these epochs, Fe lines are quite deep. The Fe abundance is high, because iron from \Nifs\ decay is now a significant contribution ($\sim 30$\%) to the Fe abundance. The feature observed near 4300\,\AA\ is still dominated by \FeIII.  The \FeIII\ feature near 5000\,\AA\ becomes broader because of the contribution of \FeII\ lines. This is reproduced reasonably well in our synthetic spectra.  

\textit{Calcium:} The synthetic spectra still reproduce well both the depth and the shape of the \CaII\,H\&K feature. At these epochs, it becomes
contaminated by \SiII\ and \CoIII\ lines in its bluer part and by \SiII\ and \FeIII\ lines in its redder part \citep[see][]{silvermanHVCa}.  The \CaII\ NIR triplet begins to appear two days after $B$ maximum, and this is reproduced in the synthetic spectra. This feature is seen much earlier in  spectroscopically normal SNe\,Ia, where it is much stronger than in SN\,1999aa even $\sim 12$\,days before maximum light \citep{Stehle2005,Mazzali2008,Tanaka2011}. Instead, in SNe\,1991T and 1999aa it only appears a few days after $B$ maximum. Calcium extends down to $v \approx 9600$\,\kms.

\textit{Silicon, Sulphur:} The shape and depth of the prominent \SiII\,6355\,\AA\ line are well replicated in the synthetic spectra. The silicon abundance is 0.25 at $v \approx 9600$\,\kms, decreasing to 0.1 at $v \approx 8900$\,\kms. 
\SII\ 5468, 5654\,\AA\ are now prominent, and increase in strength with time. Our synthetic spectra reproduce their evolution and the ratio of their depths reasonably well. The S abundance is 0.12 by mass at 
$v \approx 9600$\,\kms, decreasing to 0.05 at $v \approx 8900$\,\kms. 
 
\subsection{Post-maximum-light spectra}

\begin{figure*}
\includegraphics[trim={0 0 0 0},clip,width=0.86\textwidth]{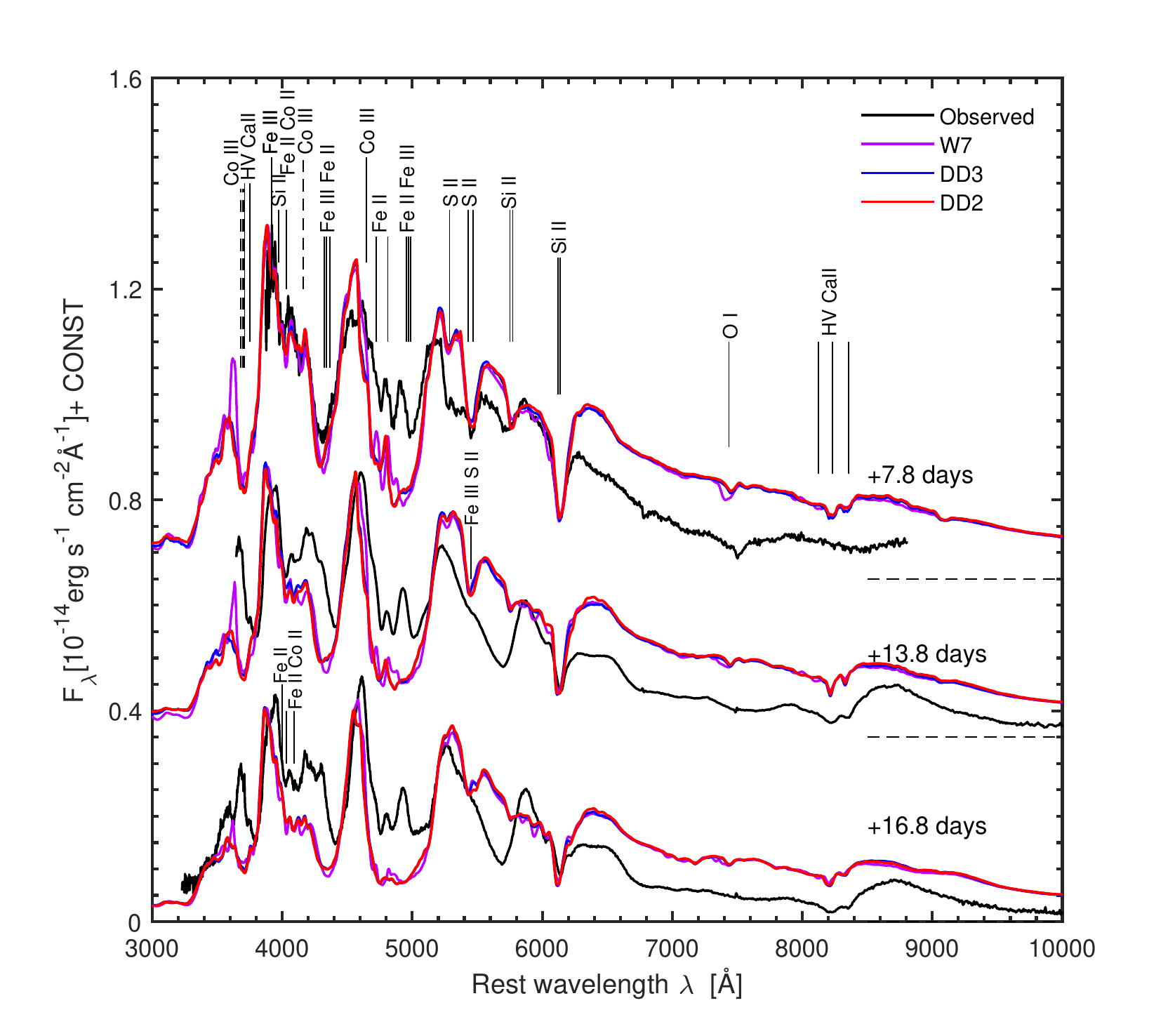}\\
\caption{Post-maximum-light spectra. Epochs are shown with reference to $B$ maximum. Features in the $+7.8$\,day spectrum near 6780\,\AA\ (very weak) and 7500\,\AA\ (weak) are telluric. The spectra have been shifted in flux by a constant value. The horizontal dashed line marks the zero flux for each epoch.}  
\label{fig11}
\end{figure*}

Fig. \ref{fig11} shows spectra ranging from 8 to 17\,days after $B$ maximum. The photospheric velocity evolves from 7150 to 4250\,\kms. At these epochs, the quality of the fits starts degrading, as the photosphere resides deep in the \Nifs-dominated region. Therefore, we do not use these epochs to infer abundances, but rather employ the nebular-phase models. Nevertheless, the synthetic spectra reproduce the observed ones sufficiently well.

\textit{Fe-group elements:} At these epochs ($\sim 30$--40\,days after the explosion), more than about 70\% of all Fe originates from the decay of \Nifs. The \FeII\ feature near 5000\,\AA\ splits into three components that are fairly reproduced in the synthetic spectra. This is the consequence of the lower degree of line blending at slower velocities.

\textit{Calcium:} The strong \CaII~H\&K line is still reproduced fairly well. The \CaII\ NIR triplet is now clearly visible, and it shows two distinct features, which are well reproduced in shape. 

\textit{Silicon, Sulphur:} The synthetic spectra still reproduce well the \SiII\,6355\,\AA\ line. \SiII\,4130\,\AA\ is now contaminated by \FeII\,4173, 4233,  4351\,\AA\ and \CoII\,4160, 4145\,\AA.
The \SII\,5468, 5654\,\AA\ lines are also contaminated by a contribution from \FeIII.  The feature near 5700\,\AA\, may be due to \NaI\,D absorption \citep{Mazzali_1990N}, or to \CoIII\ emission. We could not reproduce this feature at days $+$14 and $+$17 without major modifications to the Na ionisation structure \citep{mazzali91bg}.

\begin{figure}
\includegraphics[trim={0 0 0 0},clip,width=0.5\textwidth]{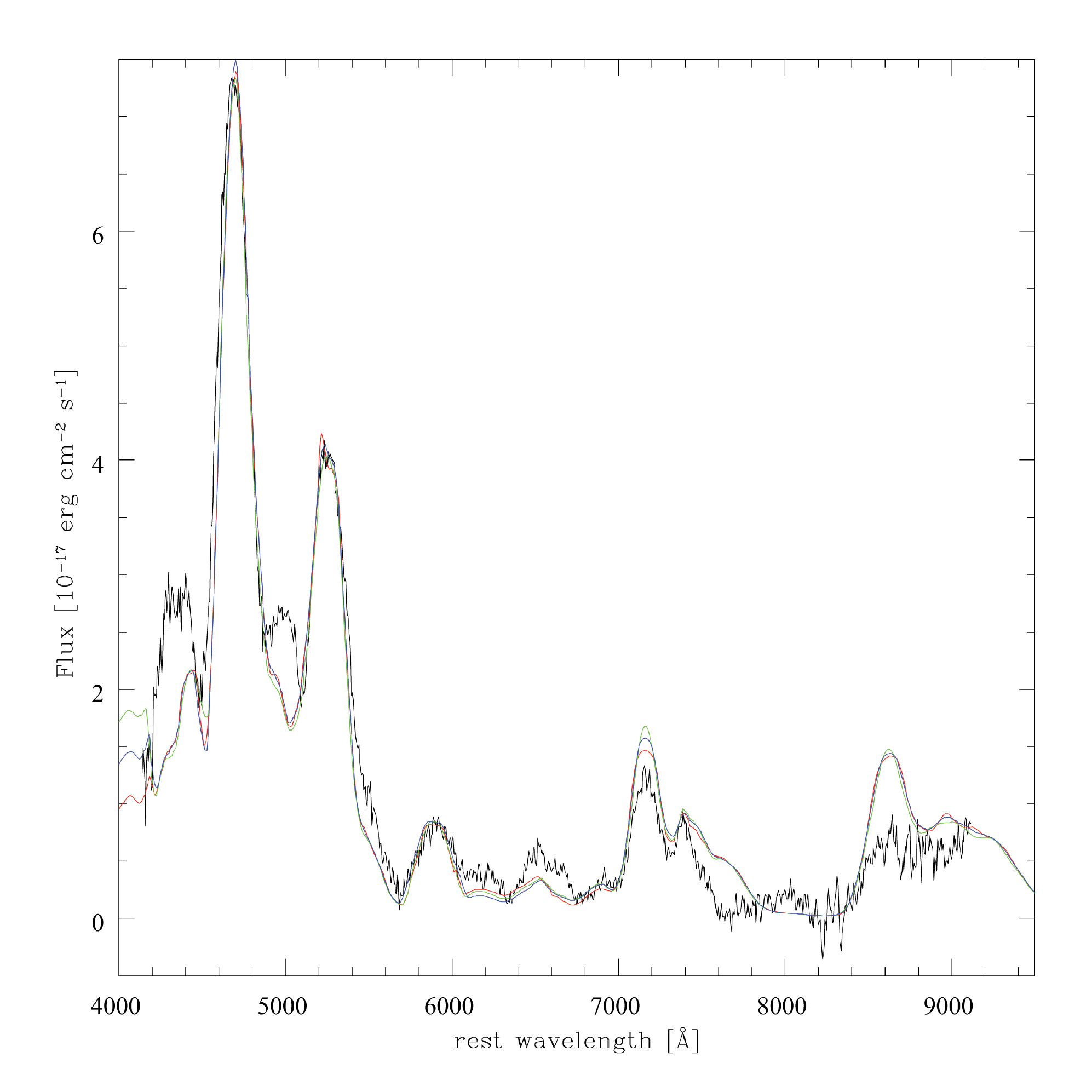}
\caption{Nebular-phase spectrum obtained on 1999 Nov. 9, corresponding to 255\, rest-frame days after $B$ maximum brightness (black). The emissions near 5200 and 7200\,\AA\ are dominated by [\FeII] lines, while the emission near 4700\,\AA\ is dominated by [\FeIII] lines. A \NaI\,D-dominated emission line is visible near 5900\,\AA, while [\NiII]\,7380 is responsible for a weak emission line. The blue, green, and red curves are the synthetic spectra obtained with DD3, DD2, and W7, respectively.}   
\label{fig12}
\end{figure}
	
\begin{figure}
\includegraphics[trim={0 0 0 0},clip,width=0.5\textwidth]{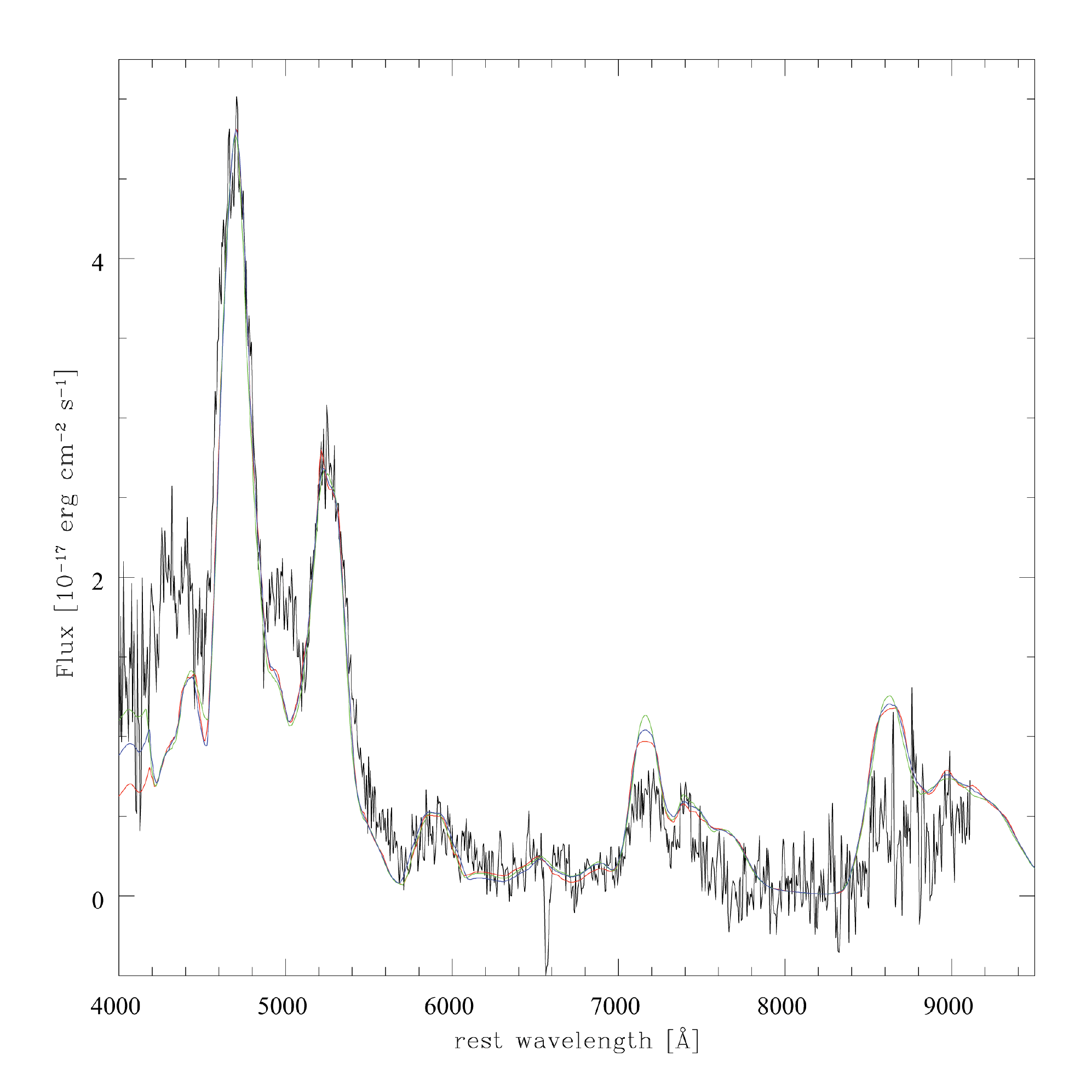}\\
\caption{Nebular-phase spectra obtained on 1999 Dec. 5, corresponding to 281 rest-frame days after $B$ maximum (black). Line identification and colour codes are similar to those in Fig. \ref{fig12}}  
\label{fig13}
\end{figure}

\section{SPECTRA IN THE NEBULAR PHASE}
\label{sec5}


Two epochs of nebular spectroscopy are available for modelling. Both were obtained with the Low-Resolution Imaging Spectrometer \citep{Oke_1995_kek_spectrometer} on the Keck-I 10\,m telescope. One spectrum was taken on 1999 Nov. 9 (exposure time 600\,s; airmass 1.01), the other on 1999 Dec. 5 (exposure time 300\,s; airmass 1.01), corresponding to 275 and 301 rest-frame days after explosion, respectively; see \citet{Silvermanetal2012} for details regarding data acquisition and reduction.

The spectra were modelled using our non-local thermodynamic equilibrium (NLTE) code, which is based on the assumptions set out by \citet{Axelrod1980}. The emission of gamma rays and positrons from a distribution of \Nifs\ is computed, and the propagation and deposition of these particles is determined using a Monte Carlo scheme as outlined first by \citet{Cappellaro1997}. Opacities $\kappa_{\gamma} = 0.027$\,cm$^2$\,g$^{-1}$ and $\kappa_{e^+} = 7$\,cm$^2$\,g$^{-1}$ are assumed in order to compute the deposition of energy. The energy that is deposited is used partly for impact-ionisation, while the rest heats the ejecta via collisional excitation. The population of excited levels is computed in NLTE. Heating is then balanced by cooling via line emission. Most emission is in forbidden lines, in particular of the elements that dominate the inner ejecta (i.e., Fe, Co, and Ni), but some is also via permitted transitions, in particular of \CaII. The ejecta are assumed to be transparent to optical radiation, so no transport is performed. As discussed by \citet[][]{mazzali2007} and others, the code can use a one-dimensional stratified density and composition, and it can be employed to explore the inner layers of an SN\,Ia and thus to complete the tomography experiment in regions that are not accessible during the early photospheric phase. 

The same three explosion models used for the early-time data are tested in the nebular phase, at both available epochs. Using the density distribution of the original models and the composition for the outer regions derived from the early-time models, we now modify the abundances in the inner layers ($v < 8000$\,\kms) in order to optimise the fits. A best fit is defined empirically, as it is impossible to match every line and not all lines carry the same weight of information, but basically we need to match both the intensity of the lines (which depends on the amount of \Nifs\ as well as of the emitting element) and their width (which traces the radial distribution of the emitting elements as well as indirectly that of \Nifs, since heating from radioactive decay must reach the emitting region). Collisional data are not perfectly known for many of the Fe lines in the optical region, so we cannot expect that all emission lines will be simultaneously reproduced. We focus therefore on reproducing the strongest emission lines. Fortunately, these include emission from both \FeIII\ (the emission near 4800\,\AA) and \FeII\  (the emission near 5200\,\AA), so we can control the ionisation of Fe, which is the dominant species in the inner ejecta at the time of the nebular spectra.  

Figs.~\ref{fig12} and \ref{fig13} show the fits to the two nebular spectra. We used the same composition at both epochs, which confirms that radioactive decay is the sole powering mechanism of the SN luminosity.

\begin{table}
\setlength{\tabcolsep}{3.5pt}
\centering                                                                  \caption{Nucleosynthetic yields and kinetic energies from the modelling using different density profiles compared to the original hydrodynamic models. Results from other SNe are also shown.}
\begin{threeparttable}
\scalebox{0.9}{
\hskip-1.0cm\begin{tabular}{llccccccc} 
\hline
  & \Nifs & $\mathrm{Fe}$\,$^a$ & $\mathrm{Ni}_{\mathrm{stable}}$ & IME\,$^b$  &O& Total burned &  $E_\mathrm{k}$ \\
  & M$_\odot$ & M$_\odot$ & M$_\odot$ & M$_\odot$ & M$_\odot$ & M$_\odot$ & $10^{51}$\,{\rm ergs} \\
\hline
1999aa (W7)  & 0.67 & 0.21  &  0.006  & 0.18  &  0.30 &1.08 & 1.2 \\
1999aa (DD2) & 0.65 & 0.29  &  0.006  & 0.20  &  0.22 &1.16 & 1.32 \\
1999aa (DD3) & 0.66 & 0.23  &  0.003  & 0.20  &  0.26 &1.12 & 1.22 \\
\hline
original W7  & 0.59 & 0.16  &  0.122  & 0.24  &  0.143 &1.24 & 1.30 \\
original DD2 & 0.69 & 0.10  &  0.054  & 0.33  &  0.066 &1.31 & 1.40 \\
original DD3 & 0.77 & 0.10  &  0.0664 & 0.25  &  0.056 &1.32 & 1.43 \\
\hline
1991T (DD3) & 0.78 & 0.15  &  0.0006  & 0.18  &  0.29 &1.09 & 1.24  \\
2003du (W7) & 0.62 & 0.18  &  0.024   & 0.26  &  0.23 &1.15 & 1.25  \\
2002bo (W7) & 0.49 & 0.27  &  0.0001  & 0.28  &  0.11 &1.27 & 1.24 \\
2004eo (W7) & 0.32 & 0.29  &  0.0005  & 0.43  &  0.3  &1.1  & 1.1 \\ 
\hline
\end{tabular}}
\begin{tablenotes}
\small      
\item $^a$All stable isotopes except for \Fefs, decay product of \Nifs 
\item $^b$ $^{28}$Si + $^{32}$S
\end{tablenotes}
\end{threeparttable}
\label{tab3}  
\end{table}                                                                                                         
The mass of \Nifs\ synthesised is $\sim 0.65$\,\Msun\ for all three models. The stable Fe mass is highest when using DD2, but it is still within the expected range of values \citep{mazzali2001}. The extra Fe seems to be located at 9000--12,000\,\kms. Stable Fe is necessary to reduce the ionisation degree and obtain a reasonable ratio of the [\FeIII] and [\FeII]-dominated features. The mass of stable Ni is quite low, and this is reflected by the weakness of the only visible Ni line, [\NiII]\,7380\,\AA. This is common to all SNe\,Ia we have studied, and suggests that little stable Ni is synthesised even in the dense innermost regions of SNe\,Ia. A moderate degree of clumping is used (filling factor $\geq 0.5$) in the \Nifs-dominated regions. Further clumping may lead to a drop in ionisation at later times, as seen in SN\,2011fe \citep{Graham_2011fe_2015} and SN\,2014J \citep{Mazzali_clumping_recomb_SN2014J}, but the available spectra of SN\,1999aa are too early and too close in time to show any change. The combined early-time and late-time studies complete the abundance tomography experiment. The main elements in the ejecta are summarised in Table \ref{tab3}, where the expected \KE\ is also shown and compared to that of the original models.

\section{ABUNDANCE TOMOGRAPHY}
\label{sec6}

\begin{figure*}
\includegraphics[trim={0 0 0 0},clip,width=0.99\textwidth]{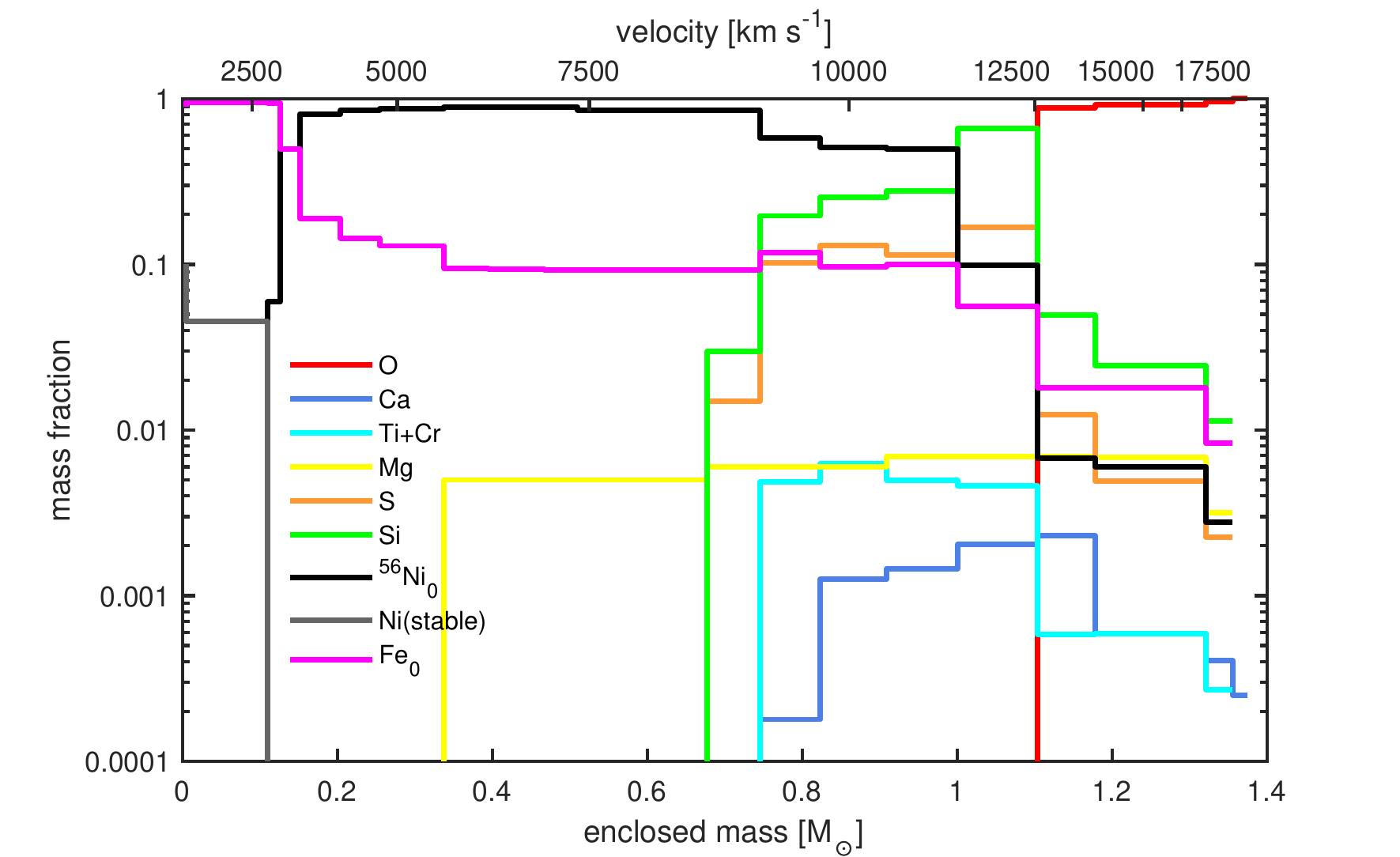}\\
\includegraphics[trim={0 0 0 0},clip,width=0.99\textwidth]{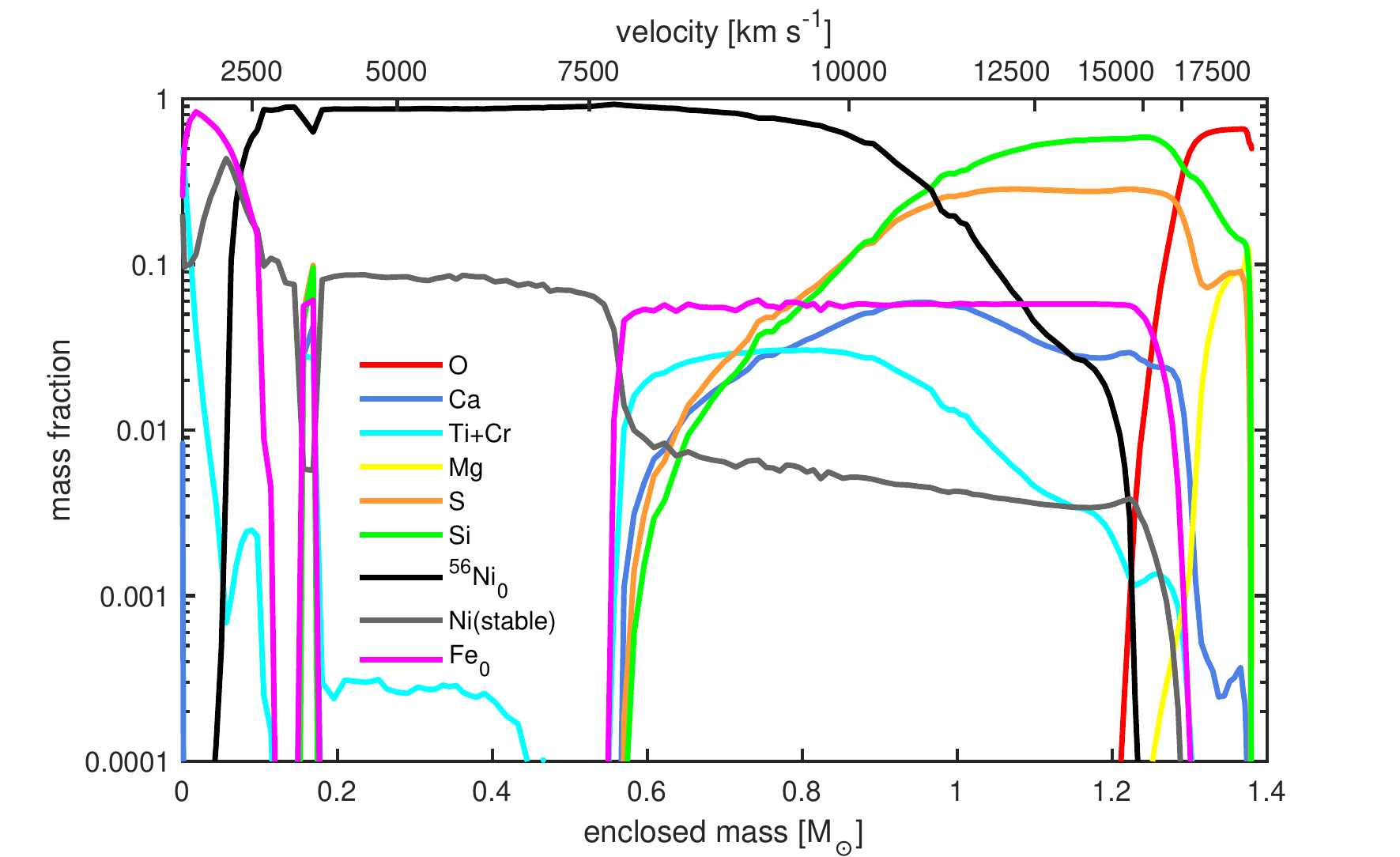}\\
\caption{Upper panel: abundances obtained from spectral models using the delayed detonation DD2 density profile. Lower panel: the original nucleosynthesis from DD2  \citep{iwamoto1999}.}  
\label{fig14}
\end{figure*}

\begin{figure*}
\includegraphics[trim={0 0 0 0},clip,width=0.99\textwidth]{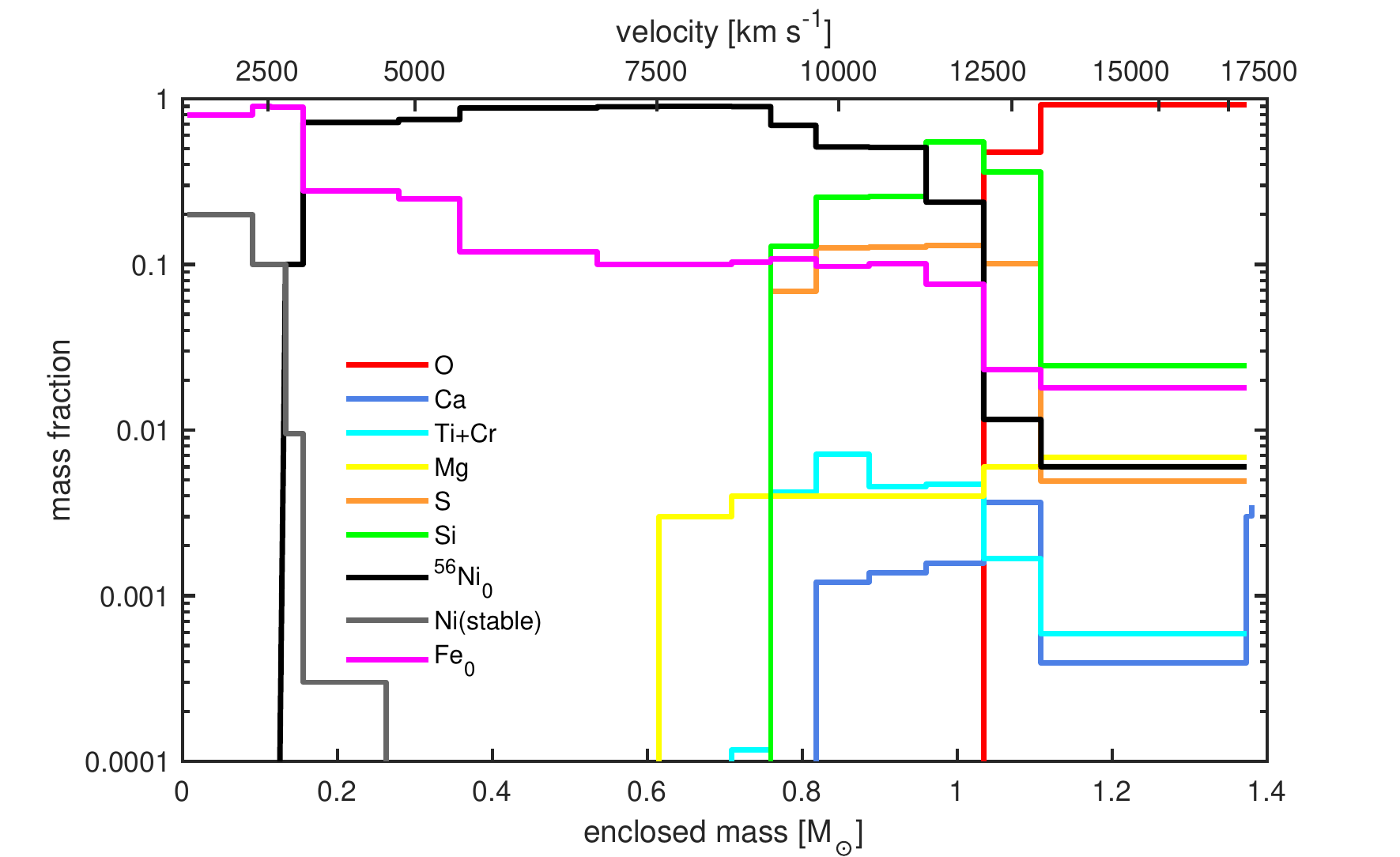}\\
\includegraphics[trim={0 0 0 0},clip,width=0.99\textwidth]{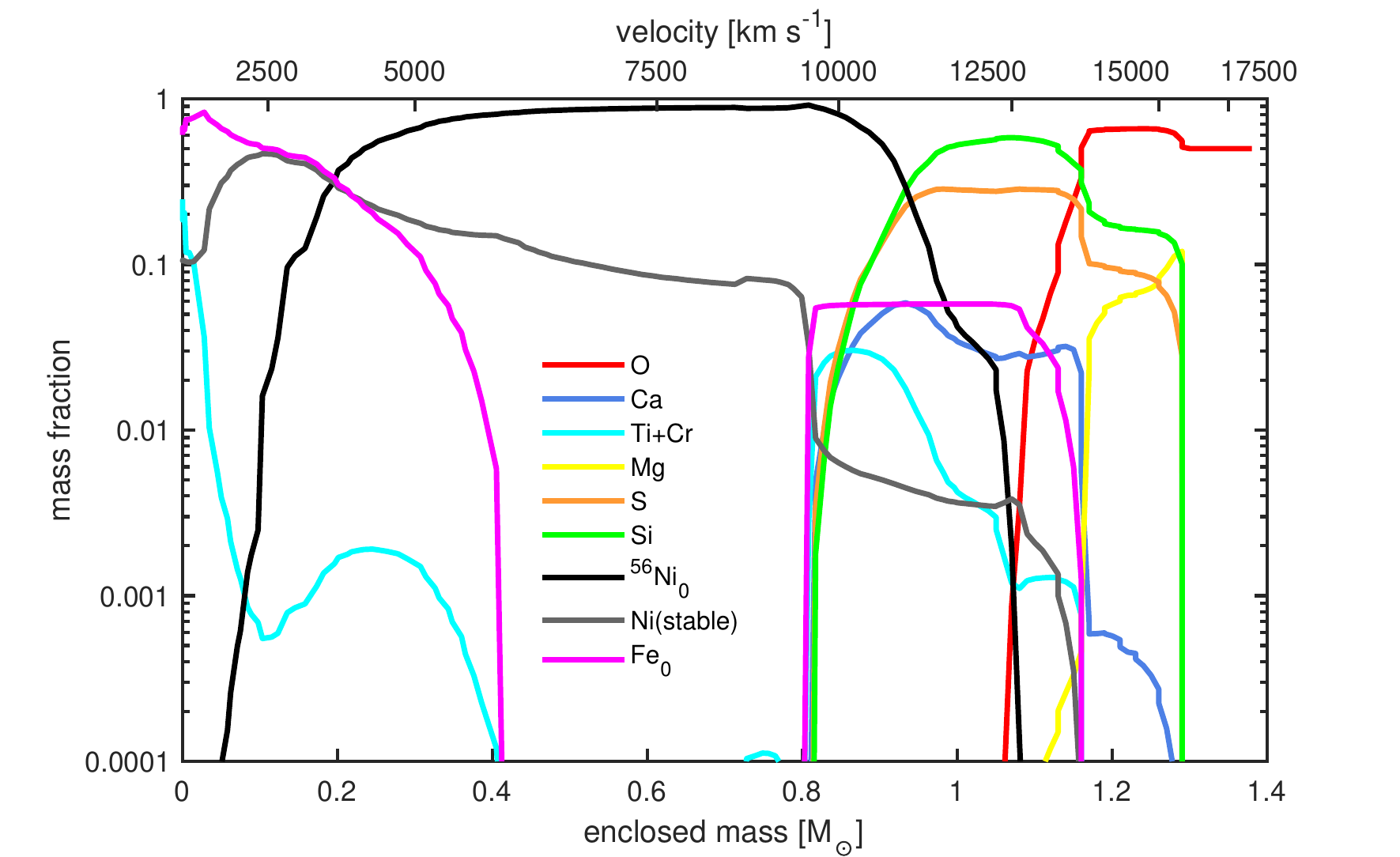}\\
\caption{Upper panel: abundances obtained from spectral models using the W7 density profile. Lower panel: the original nucleosynthesis from W7 \citep{nomoto1984}. }  
\label{fig15}
\end{figure*}

\begin{figure*}
\includegraphics[trim={0 0 0 0},clip,width=0.99\textwidth]{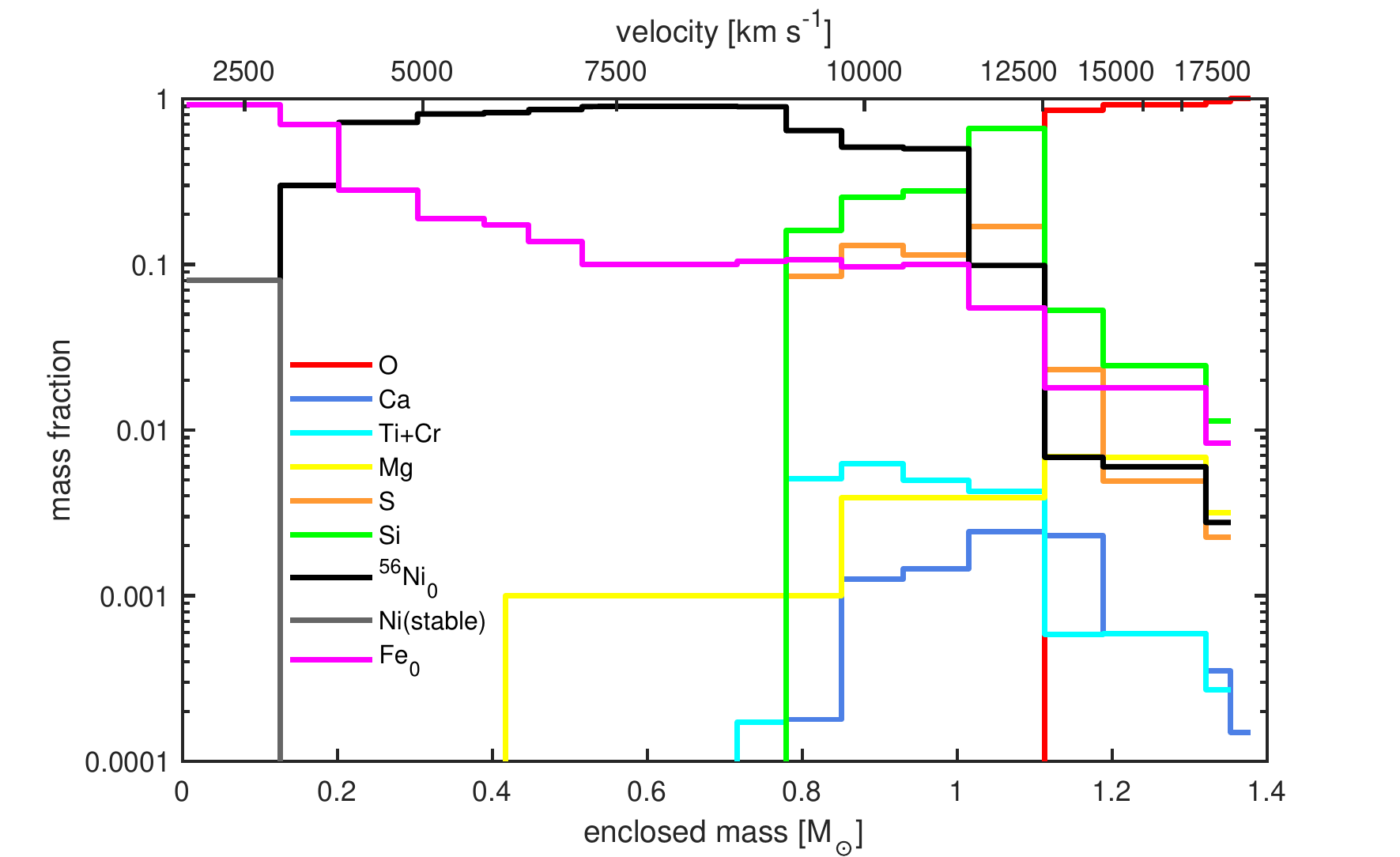}\\
\includegraphics[trim={0 0 0 0},clip,width=0.99\textwidth]{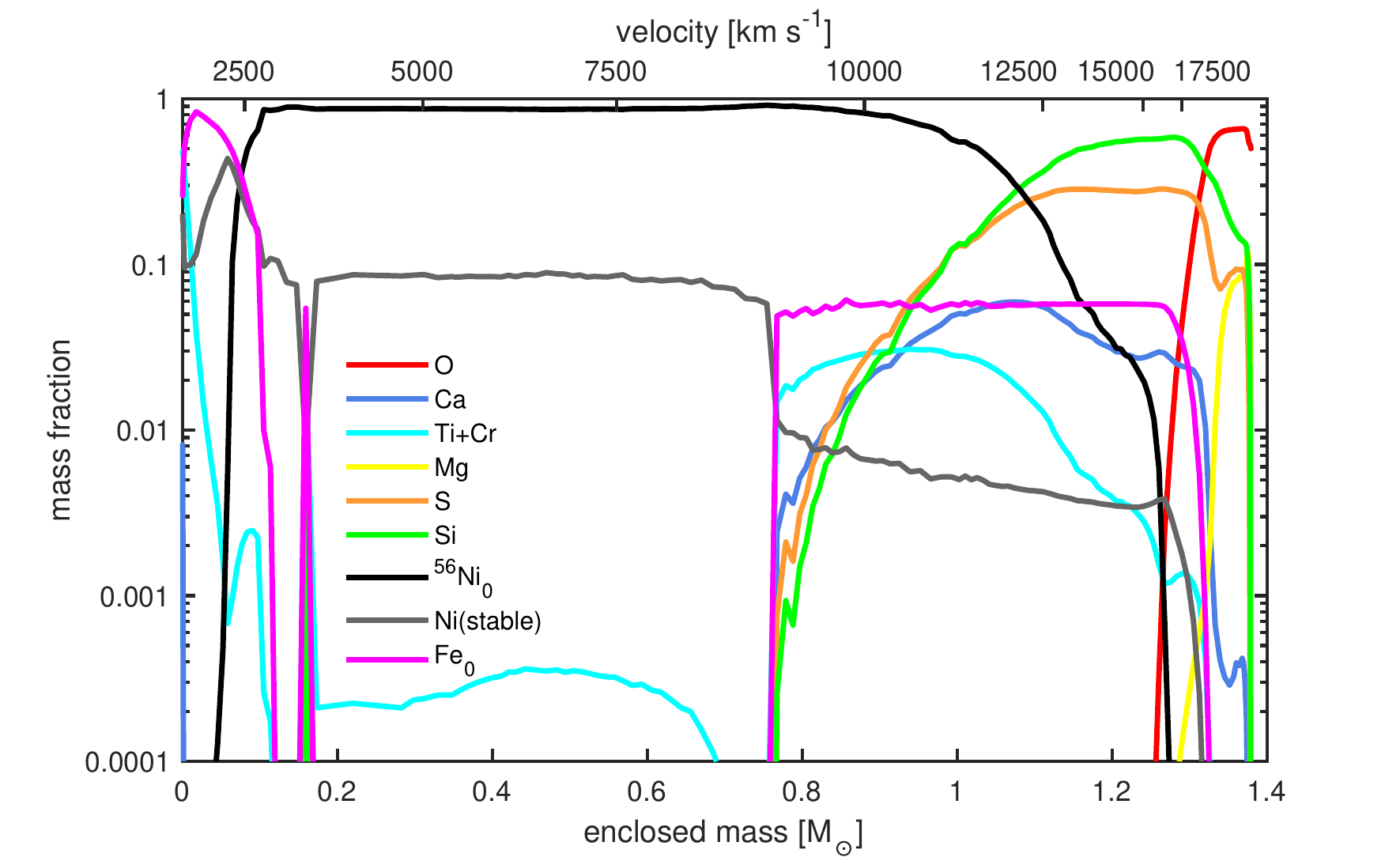}\\
\caption{Upper panel: abundances obtained from spectral models using the DD3 density profile. Lower panel: the original nucleosynthesis from DD3   \citep{iwamoto1999}.}  
\label{fig16}
\end{figure*}

The mass fractions of different elements as a function of mass and velocity for the three density profiles are shown in Figures \ref{fig14}, \ref{fig15}, and \ref{fig16}, compared to the original abundance distributions in the hydrodynamical models \citep{nomoto1984,iwamoto1999}.

The inner core, up to $v \approx 2500$\,\kms, is dominated by stable Fe with a small amount of \Nifs. Stable Fe-group elements are synthesised by electron capture in the high-density/temperature core ($\rho \geq 10^8$\,g\,cm$^{-3}$; $T \geq 5\times10^9$\,K) during the explosion, when nuclear statistical equilibrium (NSE) is attained \citep{Arnett1982,iwamoto1999,woosley}. The distribution of these elements that we derive is in general agreement with the various explosion models.

Moving outward, a \Nifs-dominated shell extends over $\sim 0.8$--1\,\Msun, out to  $v \approx 11,000$\,\kms. Practically no stable Ni is present in this region, in contrast to all explosion models, while a significant amount of stable iron is present, similar to the model prediction in the inner regions of this shell but significantly above it in regions between 3000 and 8000--9000\,\kms. This results in a larger production of stable Fe, at the expense of stable Ni, when our results are compared to the original models. 

A narrow, IME-dominated shell characterises velocities $\sim 11,000$--12,000\,\kms. The abundance of IMEs decreases sharply above this velocity. 
In the hydrodynamic models, this shell extends to higher velocities. 
The confinement of the IMEs in a narrow shell was also suggested by \citet{garavini2004} based on the velocity evolution of \SiII\,6355\,\AA.
IMEs are the result of incomplete burning, when the densities drop to $\sim 10^7\,\mathrm{g\,cm^{-3}}$. Their sudden depletion suggests a sudden drop in burning, which may be a key element to understand the structure of the progenitor and the explosion mechanism. The weakness of the IME lines in the early-time spectra of SN\,1999aa and other SN\,1991T-like SNe\,Ia is therefore an abundance effect \citep[see][]{jeffery_1992_90N_91T, filippenko19921991t}, and not only a temperature effect. The abundance of \Nifs\ is still significant in this region.

Above the IME shell, an O-rich outer layer is present. We could not conclusively determine the C abundance as no strong C features are observed. These outermost layers determine the appearance of the earliest spectra (see Fig. \ref{fig3}). Small amounts of Ca are necessary to form the \CaII\ high-velocity features (HVFs). A small abundance of stable Fe, roughly a few per cent, is necessary in order to form Fe lines at the earliest epochs. This is larger than the solar abundance.

The host-galaxy metallicity at the location of SN\,1999aa is $12 + \mathrm{log(O/H)} = 8.45$ \citep{Galbany_galaxy_metallicity}, about a factor of two below solar, suggesting that Fe at these shells is probably the result of explosive nucleosynthesis \citep[see also][]{Hachinger2013-2010jn}. In general, the presence of Fe at these shells is more consistent with DD2 and DD3 than with W7.
Only a very small amount of \Nifs\ is present in the O-rich layer, as also previously reported in other SNe\,Ia \citep{Stehle2005,Tanaka2011}. The distribution of \Nifs\ is in general consistent with the explosion models.
\vspace{-5pt}


\section{LIGHT CURVE}
\label{sec7}
\subsection{Building the bolometric light curve}
	
\begin{figure*}
\includegraphics[trim={0 0 0 0},clip,width=0.65\textwidth]{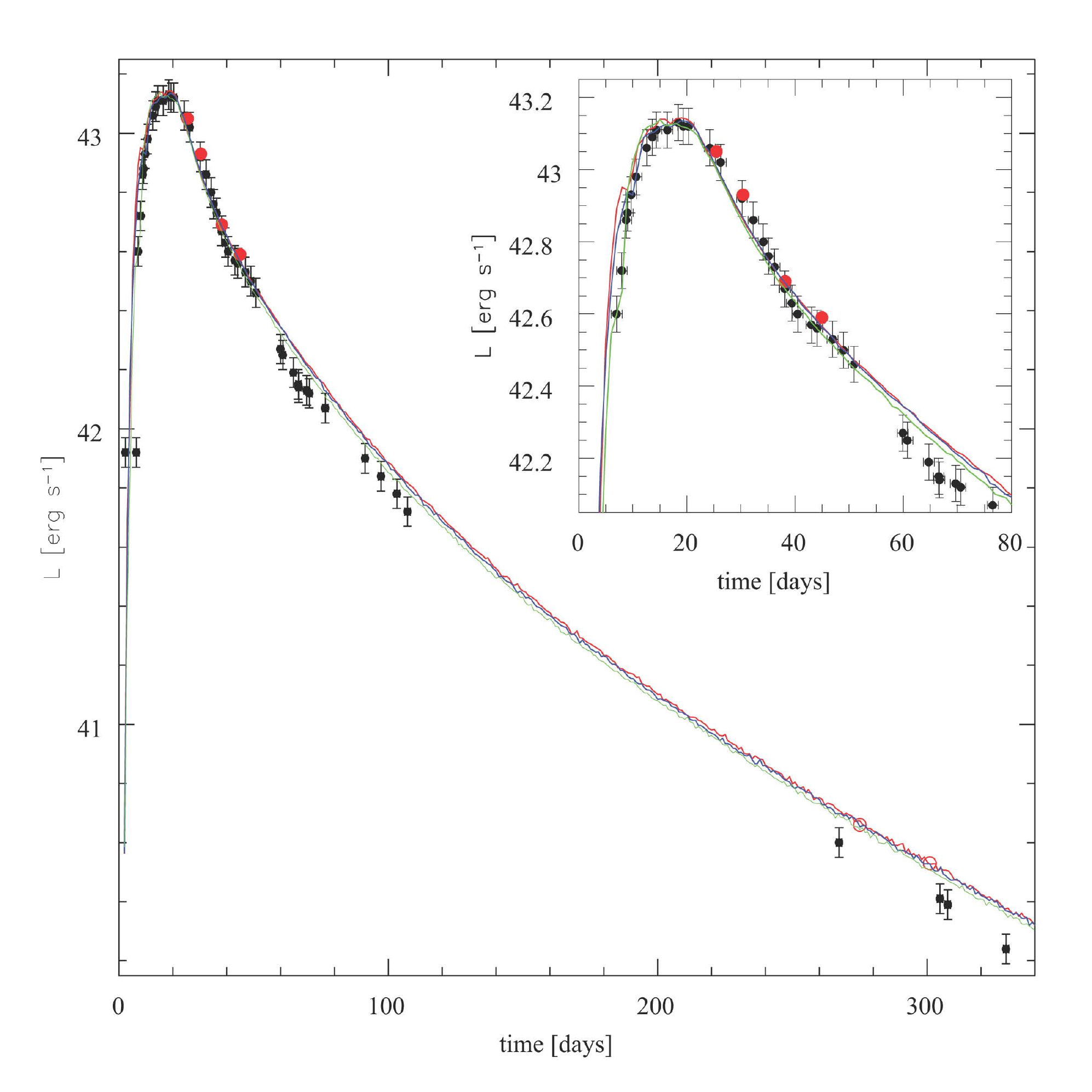}\\
\caption{The $UBVRI$ bolometric light curve of SN\,1999aa (black dots), compared to the synthetic light curves computed using the density and abundance profiles of the three explosion models: W7 (red), DD2 (green), and DD3 (blue). \cha{Red points represent luminosities at the epochs when NIR observations are available from \citet{Krisciunas_2000_photom}}. } 
\label{fig17}
\end{figure*}

We constructed a bolometric light curve of SN\,1999aa in the range 3000--10,000\,\AA. The $UBVRI$ light curves were splined with a time resolution of 1\,day, dereddened with the extinction curve of \citet{cardelli} using $E(B-V) = 0.04$\,mag \citep{Schlegel_milkywayextinction_1998} and reduced to the rest frame. Daily spectral energy distributions in the above wavelength interval were constructed using the flux zero-points of \citet{FukugitaFLUXZEROPOINTS}. For each epoch, we integrated the $U-$ to $I$-band flux after interpolating the flux between the central wavelengths of the filters, and added at the blue and red boundaries of the interval the fluxes obtained extrapolating the spectrum with a flat power law to 3000\,\AA\ and 10,000\,\AA, respectively.  The final bolometric light curve was resampled to the epochs of the actual optical observations. Since the first four measurements (i.e., prior to 1999 Feb. 13.5) are unfiltered, they have been assimilated to $V$-band fluxes and a bolometric correction was applied to them equivalent to the difference between the early bolometric magnitude and the simultaneous $V$-band magnitude. Bolometric luminosities were obtained using the luminosity distance of NGC\,2595 (63.1\,Mpc); they are shown in Fig. \ref{fig17} as black circles.

We evaluated the contribution of the NIR flux to the bolometric light curve. NIR photometry in the $J$ and $K$ bands is available at four epochs after maximum brightness \citep{Krisciunas_2000_photom}. The NIR luminosity in the range 10,000--24,000\,\AA\ was constructed following a procedure analogous to the one adopted in the optical.  Flat power laws were used to estimate the flux shortward of the $J$ band and longward of the $K$ band. Luminosities over the range 3000--24,000\,\AA\ at the four epochs when NIR observations are available are shown in Fig. \ref{fig17} as red circles. 

No UV observations are available for SN\,1999aa, so we cannot account for flux at wavelengths shorter than the Bessell $U$ filter ($\lambda < 3000$\,\AA\,). The UV should make a significant contribution only at the earliest epochs (see below). 
\vspace{-11pt}

\subsection{Modelling the bolometric light curve}

Having studied the abundance distribution for a few possible explosion models in SN\,1999aa, one way to verify the results is to test them against another observable. The light curve is one such observable. As is customary in our work, we computed synthetic bolometric light curves using the density and abundance distributions of the three models we tested. We used a Monte Carlo code that initially follows the emission and deposition of gamma rays and positrons, exactly as in the nebular spectrum calculations. The energy that is deposited is then assumed to be converted to optical photons, which are in turn transported through the ejecta using a time-dependent Monte Carlo scheme as outlined by \citet{mazzali2001}. The propagation of the optical photons is subject to an opacity. In the case of SNe\,Ia (and of all H-poor SNe), line opacity is the dominant opacity \citep{PauldrachNLTElineblocking}.
Line opacity can be parameterised based on the number of active lines in different species and the relative abundance of that species in the ejecta \citep{mazzali2001}. Photon diffusion also depends on the mass in the ejecta and their expansion (i.e., their $E_{\mathrm{k}}$). 


We computed synthetic bolometric light curves for our three explosion models. These are compared to the bolometric light curve of SN\,1999aa in Fig. \ref{fig17}. All three synthetic light curves match the observed one reasonably well. While this suggests that the models we used and the abundances we derived are credible, it is difficult to choose a best-fitting model. Although DD2 yields the closest \Nifs\ mass to the value we obtained for SN\,1999aa, the correspondence between the values we derived for the masses of the various elements and those in the original hydrodynamic calculation is not always perfect. Also, owing to the lack of early UV data, it is hard to constrain the densities in the outer layers. We can only conclude that DD2 is a reasonable model, but some modification is required.  Most likely, a specific model would have to be derived for SN\,1999aa, which may be similar to DD2 but may differ in some areas, as was the case for SN\,2011fe \citep{mazzali_2014_2011fe,mazzali20152011fe}.
\vspace{-15pt}

\section{DISCUSSION}
\label{sec8}

Our synthetic spectra show reasonably good fits to the observed ones for the three density profiles used, with only small differences between them.
For example, the \SiII\,6355\,\AA\ feature (Fig.~\ref{fig4}), the \OI\,7744\,\AA\ line (Fig.~\ref{fig8}), and the \FeII\ lines near 5000\,\AA\ (Fig.~\ref{fig10}) are better reproduced with the DD2 and DD3 density profiles than with W7. However, these differences are marginal, and based on this criterion alone it is difficult to select a best-fit model.  

The yields of the most important elements or groups of elements are recapped in Table \ref{tab3}. From these yields we computed the expected kinetic energy yield for each model using the formula 
\setlength{\abovedisplayskip}{5pt}
\setlength{\belowdisplayskip}{5pt}
\begin{multline}
    E_{k}=[1.56\,M(\ce{^{56}Ni}) + 1.47\,M(\mathrm{Fe}) +    1.24\,M(\mathrm{IME}) - E_{\mathrm{bind}}]\\ \times 10^{51}{\rm erg}
\end{multline}
\citep{woosley}, where $E_{\mathrm{bind}} = 0.46\,10^{51}$\,erg is the binding energy of the white dwarf. Results are given in Table~\ref{tab3}. 
The values we obtain are slightly smaller than the original models. The difference may be explained by the weak burning at the outer shells. A similar behaviour was seen in SN\,1991T \citep{Sasdelli2014}.
\setlength{\parskip}{0.5em}

\begin{figure*}
\includegraphics[trim={0 0 0 0},clip,width=0.80\textwidth]{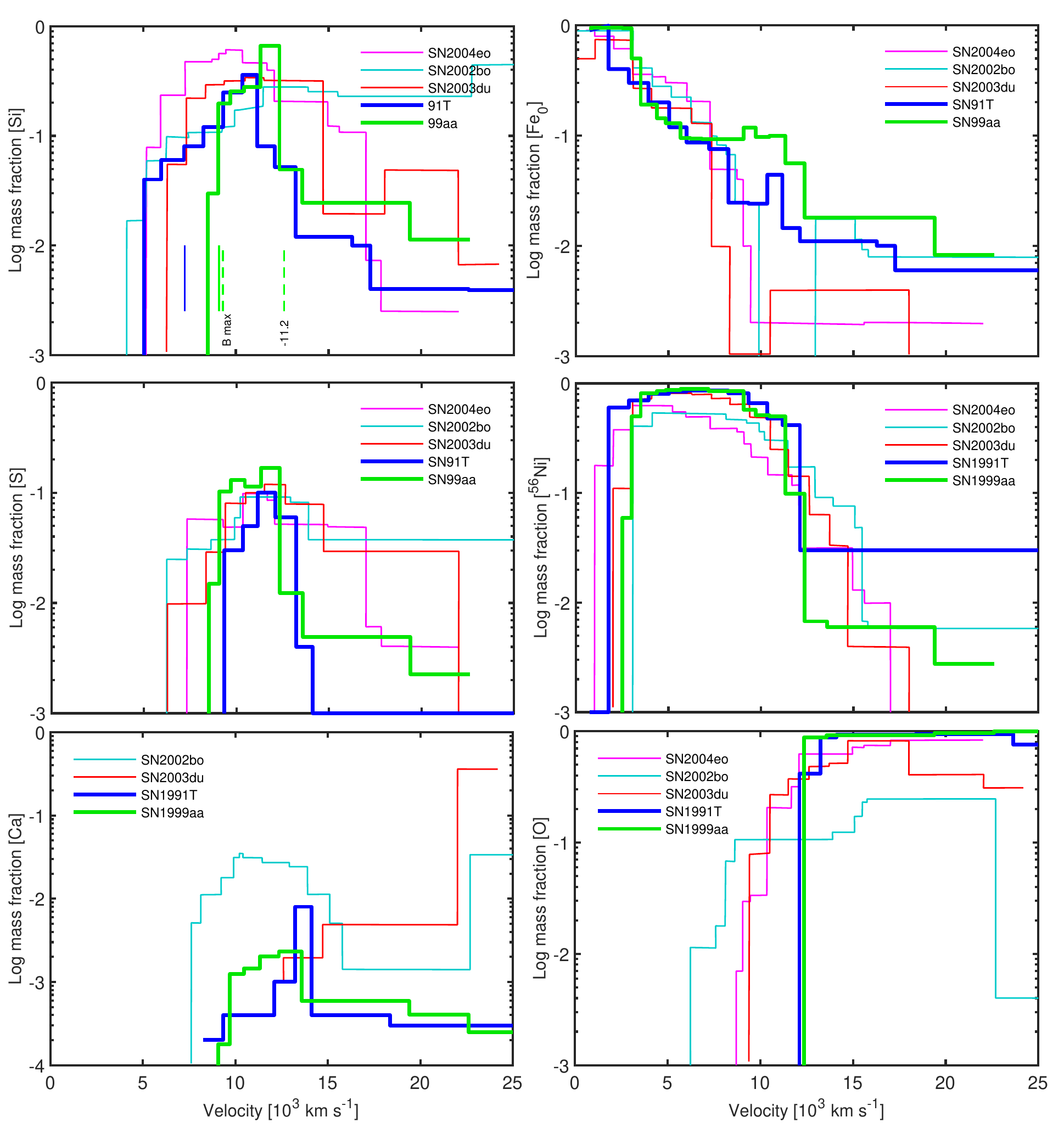}\\
\caption{The distribution of the most important elements in SN\,99aa, SN\,1991T, and some spectroscopically normal SNe\,Ia. Left-hand side, top to bottom: Si, S, and Ca. Right-hand side, top to bottom: stable Fe, \Nifs, O. SNe\,1999aa and 1991T have similar stratification properties: a more complete dominance of \Nifs\ in the inner layers (2000--10,000\,\kms), a narrow IME shell peaking near 11,000\,\kms\ but terminating sharply above $\sim 12,000$\,\kms, a larger prevalence of oxygen in the outer layers, suggesting less burning in these regions. The dashed lines in the first panel show $v_{\mathrm{ph}}$ at days $-$11.2 and at $B$ maximum light. The continuous lines show the position of $v_{\mathrm{ph}}$ at the epochs when \FeII\ lines start to appear in SN\,1999aa (green) and SN\,1991T (blue).}  
\label{fig18}
\end{figure*}

In Fig. \ref{fig18} we compare the abundance distribution of SN\,1999aa with those of three spectroscopically normal SNe\,Ia --- SN\,2003du \citep[\Dm = 1.13\,mag;][]{Tanaka2011}, SN\,2002bo \citep[\Dm = 1.13\,mag;][]{Stehle2005}, and SN\,2004eo \citep[\Dm = 1.46\,mag;][]{Mazzali2008} --- and with that of SN\,1991T \citep[\Dm = $0.94 \pm 0.05$\,mag;][]{Phillips_1999_deltam_1991T}. The plots show a striking similarity between the internal composition of SNe\,1999aa and 1991T, but clear differences with respect to spectroscopically normal SNe\,Ia.


Most significantly, although IMEs reach a high abundance in a shell at $\sim 11,000$\,\kms, the IME-dominated shell is very narrow, and therefore has little mass. At the outer edge, unlike in normal SNe\,Ia, the abundance of IMEs drops very sharply at $v \approx 12,000$\,\kms, above which oxygen dominates. 

This suggests that the weakness or absence of \SiII\ and \SII\ features in the earliest spectra of SN\,1991T-like SNe\,Ia is not only an ionisation effect but also the result of a low abundance \citep{mazzali1995}. In these peculiar SNe the IME abundance in the outermost layers is very small, and therefore the spectra start looking like those of normal SNe\,Ia at a later time. 

\begin{figure}
\includegraphics[trim={0 0 0 0},clip,width=0.45\textwidth]{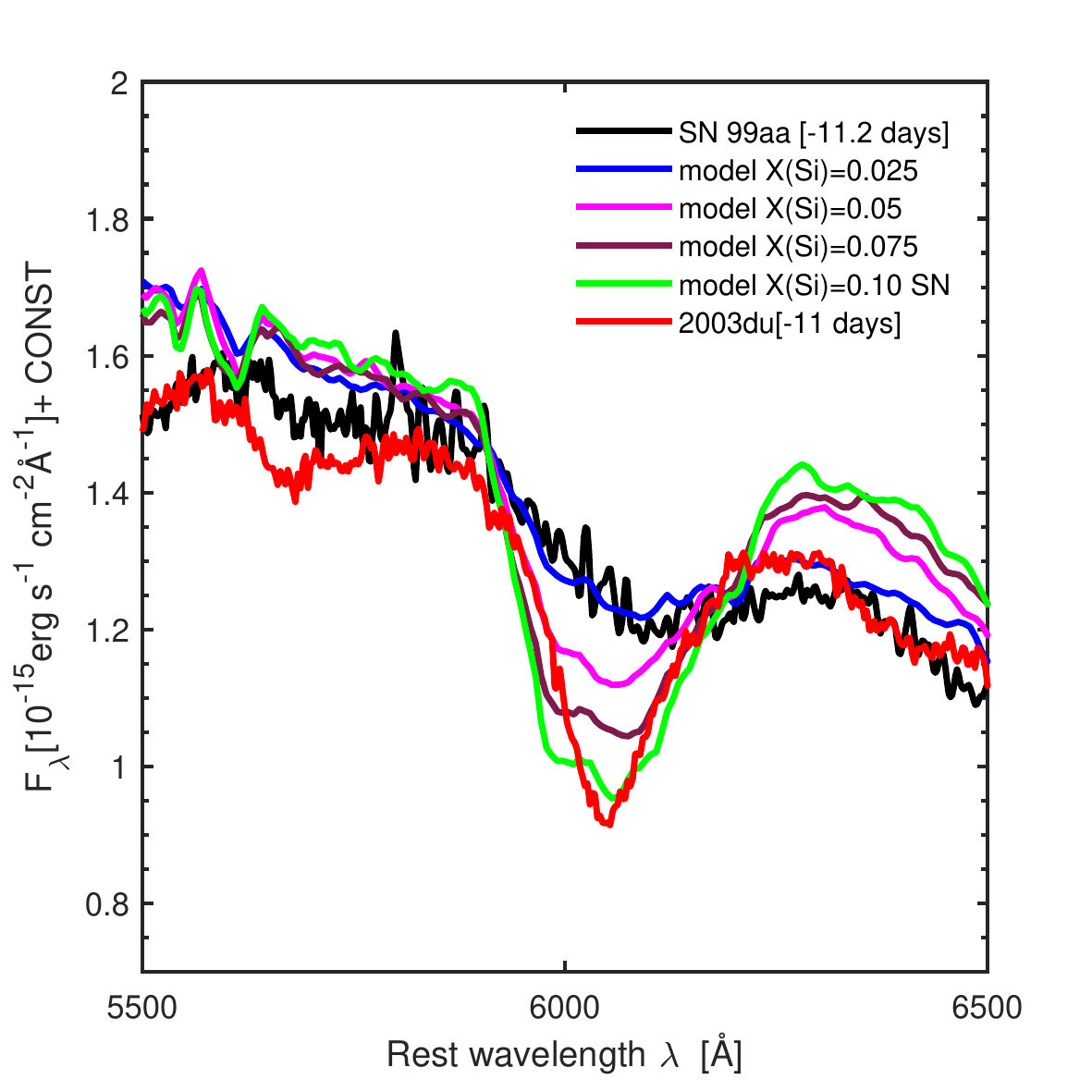}\\
\caption{The effect of changing the Si abundance in the outermost layers.} 
\label{fig19}
\end{figure}

In order to check the effect of the Si abundance on the spectra, we computed synthetic spectra using increasing quantities of Si at $v \approx 12,600$\,\kms, at the expense of O, while keeping $L_{\mathrm{Bol}}$ and $v_{\mathrm{ph}}$ unchanged (Fig.\,\ref{fig19}). The \SiII\,6355\,\AA\ line in SN\,1999aa is well reproduced with a Si abundance of $\sim 0.025$. As the Si abundance increases the line gets stronger, and it matches the spectrum of SN\,2003du when the abundance is $\sim 0.1$ at high velocities, which is comparable to the abundances reported in SN\,2002bo \citep{Stehle2005} and SN\,2004eo \citep{Mazzali2008}. The abundance derived by \citet{Tanaka2011} for SN\,2003du is even higher ($\sim 0.3$ at $v \approx$ 10,500--15,000\,\kms).  

\begin{figure}
\includegraphics[trim={0 0 0 0},clip,width=0.42
\textwidth]{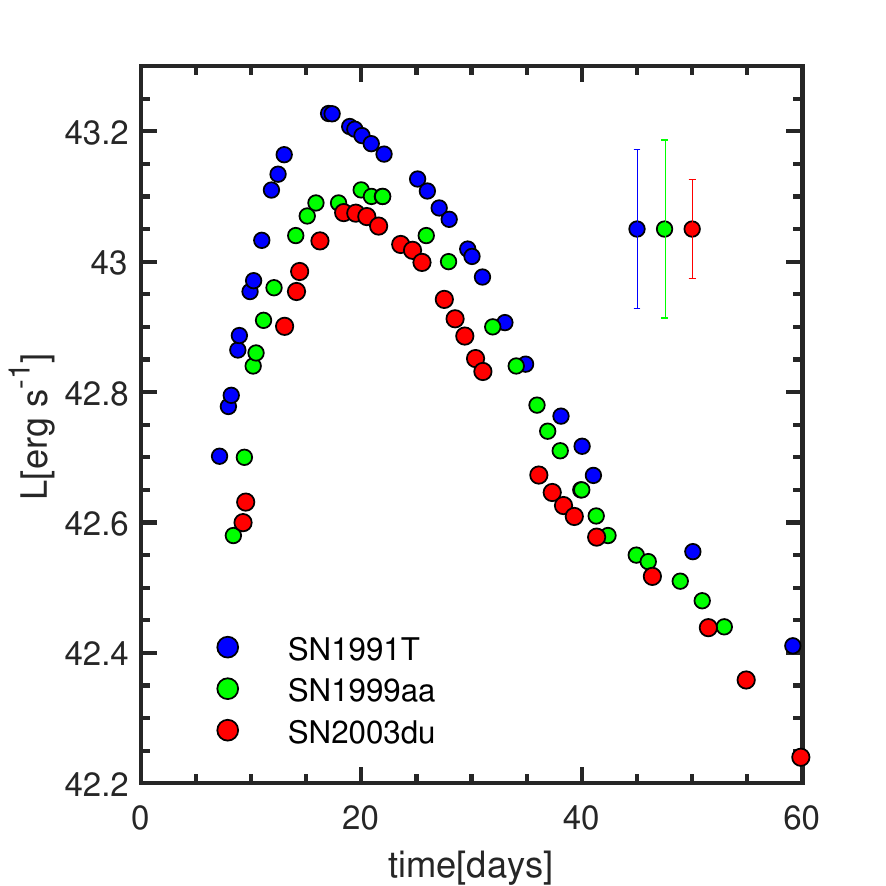} \caption{The bolometric light curve of SN\,1999aa (green) compared to those of SN\,1991T \citep[][blue]{Sasdelli2014} and SN\,2003du \citep[][red]{stanishev2003du}. The UV contribution was removed from the light curve of SN\,2003du to allow a proper comparison, as no UV information is available for the other two SNe. \cha{Error bars for SN\,1999aa and SN\,1991T represent statistical errors due to distance moduli uncertainties $\delta\mathrm{\mu}$, to the host galaxies using the Tully-Fisher relation: $\delta\mathrm{\mu}$[99aa]=0.34, $\delta\mu$[91T]= 0.30. The error bar for SN\,2003du is calculated taking $\delta\mu$[03du]=0.19 \citep{stanishev2003du}}. }  
\label{fig20} 
\end{figure}

Although the spectroscopic properties of SN\,1999aa suggest that it is physically intermediate between SN\,1991T and normal SNe\,Ia, its photometric properties do not. Our modelling shows that the amount of \Nifs\ synthesised in SN\,1999aa ($\sim 0.65$\,\Msun) is less than in SN\,1991T \citep[$\sim 0.78$\,Msun;][]{Sasdelli2014},   suggesting that SN\,1999aa should be less luminous than SN\,1991T (see Fig. \ref{fig20}). 
However, SN\,1999aa was a slower decliner than SN\,1991T. SN\,1999aa has estimated \Dm\ values ranging from 0.75\,mag \citep{Krisciunas_2000_photom} to 0.85\,mag \citep{Jhaetal2006}, which may be taken to imply that it was actually more luminous than SN\,1991T (\Dm = 0.94\,mag). 

However, a comparison of the bolometric light curves of the two SNe shows that relying on a \Dm\ alone would be misleading. The light curve of SN\,1991T is brighter throughout, as it should be based on the \Nifs\ mass. However, it peaks much earlier than that of SN\,1999aa. This is because the  \Nifs\ abundance in the outer layers of SN\,1991T is larger than in SN\,1999aa, causing a faster rise to a very luminous maximum (see Fig.\ref{fig18}). The luminous phase is then sustained by the larger \Nifs\ mass, but the contrast between the nominal luminosity at peak and that 15\,days later is larger than in SN\,1999aa, which reaches maximum brightness later. This may mean that \Dm is not valid for the SN\,1991T class \citep[see also][]{pinto_eastman_LC,woosley,scalzo_2012}, and it was also suggested for objects at the faint end of the luminosity-width relation \citep{Ashall_2018_SN2007on_SN2011iv}. \cha{ On the other hand, SN\,2003du and SN\,1999aa, reach peak luminosities that differ by only $\log{(L)}\lesssim$~0.05, even though they have different decline rates and different spectroscopic properties. Despite the  distance uncertainties, this result can be taken to confirm that both of these events synthesize a similar mass of \Nifs\ as suggested from our spectral modeling ($\sim$~0.62-0.65 \Msun, see Tab \ref{tab3}) .}


\begin{figure}
\includegraphics[trim={0 0 0 0},clip,width=0.45\textwidth]{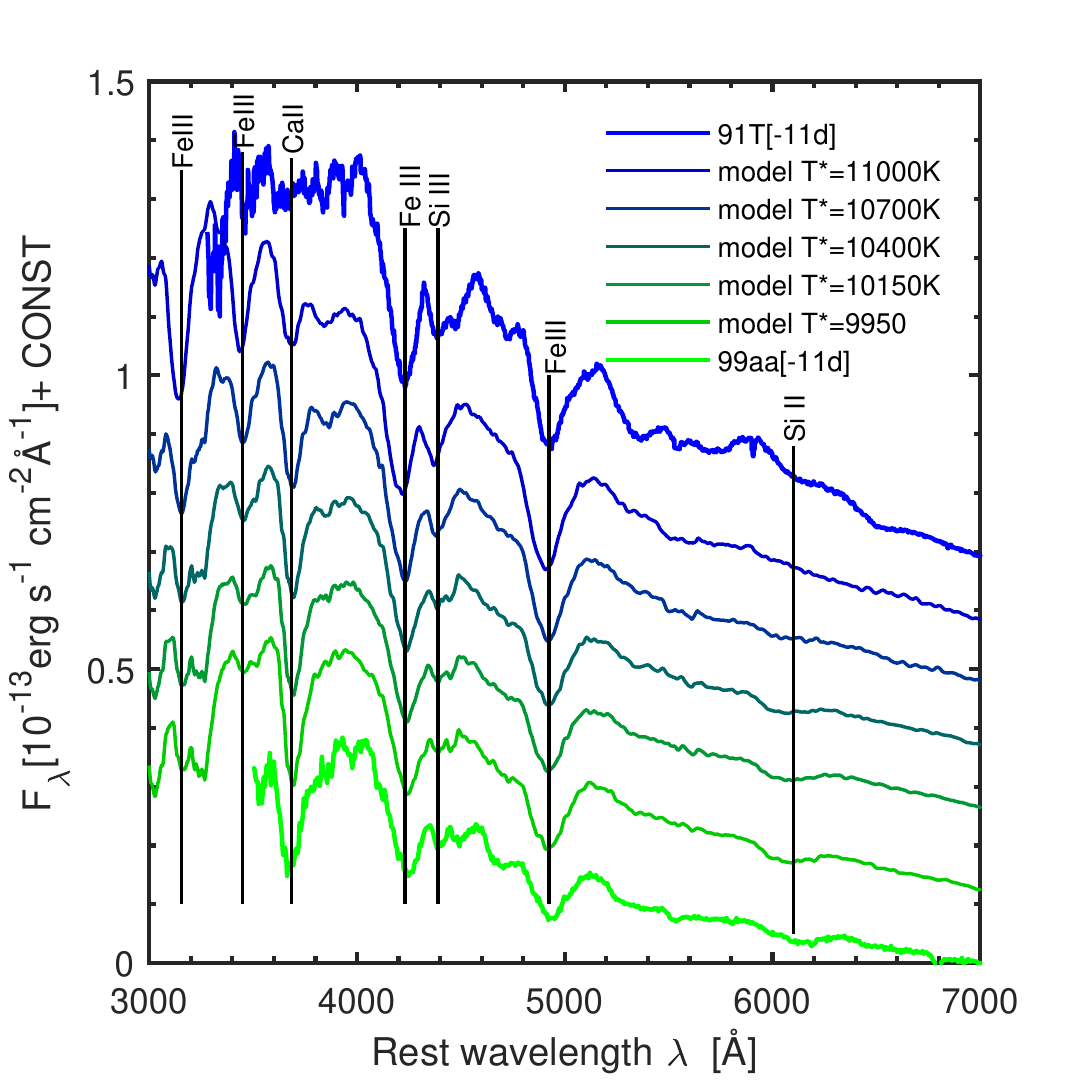}\\
\caption{Early-time spectra of SNe\,1999aa and 1991T compared to synthetic spectra computed for increasing luminosity but the same composition. As the luminosity increases the spectra morph from looking like SN\,1999aa to looking like SN\,1991T.}  
\label{fig21}
\end{figure}

Even though the abundance distributions in SNe\,1999aa and 1991T are similar, their spectroscopic evolution shows differences. These can be explained by the difference in luminosity between the two SNe. We computed synthetic spectra at day $-$11 starting from the model that matches SN\,1999aa and progressively increased the luminosity (Fig.\ref{fig21}). As the luminosity increases, the spectrum changes, until it finally starts resembling that of SN\,1991T: the \SiII\,6355\,\AA\ line becomes weaker, and so does \CaII\,H\&K. The same is true for the \FeIII\ features observed near 3200, 3500, 4200, and 4900\,\AA.

SNe\,Ia exhibit very similar spectroscopic properties beyond maximum brightness. Therefore, an explosion-progenitor scenario that can explain the complete spectroscopic sequence should be one that allows variations only in the outer layers. The sudden depletion of the IMEs in the outer shells of SN\,1999aa is not easy to explain within the framework of conventional delayed-detonation explosion models \citep{khokhlov91DD,iwamoto1999}.  
One possible explanation may be an explosion that initially proceeds very efficiently but then suddenly stops, leaving an only weakly burned outer layer. One such class of models is pulsation-driven detonations 
\citep{ivanova_PDD_1974,Khokhlov1991_DD_PDD, Hoeflich1995ApJ-pulsation_delayed_detonations}.
In these configurations, the progenitor is characterised by an outer layer with low density, which could be the result of the pre-expansion of a white dwarf that has gone through an initial failed burning phase, or to a binary merger. This results in a steep density gradient and may cause IMEs to be confined in a relatively narrow velocity range. However, these models predict no burning in the outermost layers, and therefore the presence of a copious amount of C \citep{baron_2008_PRD_showing_C}, which is not observed in SN\,1999aa or SN\,1991T. Additionally, simulations of these models show IME lines at very early times, and do not resemble the spectra of SN\,1991T-like SNe\,Ia \citep{dessart2014pdd}. Furthermore, three-dimensional versions of these models exhibit a large degree of mixing and cannot explain the stratification seen in SN\,1999aa  \citep{Plewa2004,kasenGCD2005,BRAVOPRD1,BRAVOPRD2}.
 
In general, none of the currently available models can explain the entire spectroscopic properties of SNe\,Ia over a large range of luminosities. Nevertheless, the pulsation-driven scenario remains interesting for SN\,1991T-like SNe because it only affects the outer ejecta. Based on our current knowledge, this particular scenario should only kick in when \Nifs\ production is very high.
\vspace{-5pt}

\section{CONCLUSIONS}
\label{sec9}

We have modelled a series of optical spectra of the peculiar slow decliner SN\,1999aa, from $-$12 to $+\sim 300$\,days from $B$ maximum to infer the composition layering of its ejecta.
Three different density profiles were used --- the fast deflagration W7 and two delayed detonation models, DD2 and DD3. We have compared our results with spectroscopically normal events as well as with SN\,1991T. 

Our main results can be summarised as follows.
\begin{itemize}
\setlength{\itemsep}{5pt}
\item All three density profiles yield synthetic spectra similar to the observed ones and follow their evolution. In particular, an \FeIII-dominated early-time spectrum with shallow IME lines, typical of the SN\,1991T class, is reproduced.
\item The internal composition of SN\,1999aa is dominated by neutron-rich iron-peak elements, as in normal SNe\,Ia. This is followed by a \Nifs\ shell (mass $\approx 0.65$\,\Msun). Above this lies a narrow IME shell which is sharply separated from the outer, O-dominated shell. 
\item The confinement of IMEs to a narrow velocity range and their depletion in the 
outermost layers indicates a sudden shift from a regime of strong burning to one of weak incomplete burning. This behaviour is remarkably similar to that of SN\,1991T, but is not observed in normal SNe\,Ia.  Therefore, it is reasonable to conclude that SNe 1999aa and 1991T share a similar explosion mechanism, despite their somewhat different luminosities.
\item The observed stratification may be the result of sharp density gradients in the outer shells of the progenitor.
\item The spectroscopic path from normal SNe\,Ia to the brightest peculiar events cannot be explained solely by a luminosity/temperature sequence. It should involve composition layering differences suggesting variations either in the density structure of the progenitor white dwarf at the outer layers or in details of the explosion. 
\item Within the SN\,1991T class, IME confinement coupled with differences in luminosity (i.e., \Nifs\ production) may explain the observed spectra. 
\end{itemize}

\vspace{-25pt}
\section*{ACKNOWLEDGMENTS}

This work has used data collected at the Italian Telescopio Nazionale Galileo (TNG), which is operated on the island of La Palma by the Fundacion Galileo Galilei of the INAF (Istituto Nazionale di Astrofisica) at the Spanish Observatorio del Roque de los Muchachos of the Instituto de Astrofisica de Canarias. \cha{The authors thank the anonymous referee for his well thought comments that improved the quality of this study}.  CA thanks Roger Hajjar and Sami Dib for discussions, and Amy Sidaoui for English review.

A.V.F. is grateful for financial support provided by the U.S. National Science 
Foundation, the Christopher R. Redlich Fund, and many individual donors.
\cha{This research has made use of the NASA/IPAC Extragalactic Database (NED) which is operated by the Jet Propulsion Laboratory, California Institute of Technology, under contract with the National Aeronautics and Space Administration.}

Some of the data presented herein were obtained at the W.~M. Keck
Observatory, which is operated as a scientific partnership among the
California Institute of Technology, the University of California, and
NASA; the observatory was made possible by the generous financial
support of the W.~M. Keck Foundation. The Kast spectrograph on the
Shane 3\,m telescope at Lick Observatory was made possible through
a gift from William and Marina Kast. Research at Lick Observatory is        
partially supported by a generous gift from Google. We thank the staffs
at the various observatories where data were obtained.

\vspace{-15pt}

\section*{DATA AVAILABILITY}

The spectroscopic data used in this article are available at the Weizmann Interactive Supernova Data Repository (WISeREP) \citep{wiserep}.

\vspace{-15pt}




\bibliographystyle{mnras}
\bibliography{references}









\bsp	
\label{lastpage}
\end{document}